\documentclass[12pt,preprint]{aastex}

\slugcomment{To appear in the Astrophysical Journal}

\shorttitle{Flaring in the young Sun}
\shortauthors{Feigelson et al.}

\begin{document}

\title{Magnetic flaring in the pre-main sequence Sun and \\
implications for the early solar system}

\author{Eric D. Feigelson, Gordon P. Garmire}
\affil{Department of Astronomy \& Astrophysics, Pennsylvania
State University, \\ University Park PA 16802} 
\email{edf@astro.psu.edu}
\and
\author{Steven H.  Pravdo}
\affil{Jet Propulsion Laboratory, MS 306-438, 4800 Oak
Grove Drive, \\ Pasadena CA 91109}

\begin{abstract}

To address the role of energetic processes in the solar nebula, we
provide a detailed characterization of magnetic flaring in stellar
analogs of the pre-main sequence Sun based on two 0.5-day observations
of analogs of the young Sun in the Orion Nebula Cluster obtained with
the {\it Chandra X-ray Observatory}. The sample consists of 43 stars
with masses 0.7-1.4 M$_\odot$ and ages $<0.3$ to $\simeq 10$ My.  We
find the X-ray luminosities measured in the $0.5-8$ keV band are
strongly elevated over main sequence levels with average $<\log L_x> =
30.3$ erg s$^{-1}$ and $<\log L_x/L_\star> = -3.9$.  The X-ray emission
is strongly variable within our exposures in nearly all solar analogs;
about 30 flares with $29.0 < \log L_x(peak) < 31.5$ erg s$^{-1}$ on
timescales from $0.5$ to $>12$ hours are seen during the Chandra
observations. Analogs of the $\leq 1$ My old pre-main sequence Sun
exhibited X-ray flares that are $10^{1.5}$ times more powerful and
$10^{2.5}$ times more frequent than the most powerful flares seen on
the contemporary Sun.  

Radio observations indicate that acceleration of particles to relativistic
energies is efficient in young stellar flares.  Extrapolating the solar
relationship between X-ray luminosity and proton fluence, we infer that
the young Sun exhibited a $10^5$-fold enhancement in energetic protons
compared to contemporary levels.  Unless the flare geometries are
unfavorable, this inferred proton flux on the disk is sufficient to
produce the observed meteoritic abundances of several important
short-lived radioactive isotopes.  Our study thus strengthens the
astronomical foundation for local proton spallation models of isotopic
anomalies in carbonaceous chondritic meteorites.  The radiation,
particles and shocks produced by the magnetic reconnection flares seen
with $Chandra$ may also have flash melted meteoritic chondrules and
produced excess $^{21}$Ne seen in meteoritic grains.

\end{abstract}


\section{Introduction}

A large body of meteoritic evidence indicates that violent high-energy
processes occurred in or near the solar nebula during the formation of
the solar system.  The prevalence of chondrules requires sudden heating
of solid material to $T\simeq 2000$ K, likely produced by shocks, lightning
or radiation in the solar nebula \citep[see reviews in][]
{Hewins96, Jones00}. Anomalous abundances of elemental isotopes are
seen in chondrules and inclusions of the most pristine carbonaceous
chondrites \citep[see review by][]{Goswami00}.  Most puzzling are the
high abundances of daughter products of short-lived nuclides like
$^{41}$Ca, $^{26}$Al and $^{53}$Mn.  These nuclides are often
attributed to the injection of freshly produced nuclear products into
the nebula from stellar sources:  a supernova remnant or winds from an
asymptotic giant branch (AGB) or Wolf-Rayet star \citep{Cameron77,
Wasserburg94, Arnould97}.  

An alternative explanation involves the production of unusual nuclear
species within the nebula by spallation reactions from energetic (MeV)
protons and ions originating in magnetic reconnection flares.
Energetic particles are currently produced in the solar system by
flares above the Sun's surface, but with fluences far below those
required by local irradiation models of short-lived meteoritic
nuclides.  In early calculations \citep[e.g.][] {Hoyle62, Clayton77,
Lee78}, local irradiation models had difficulty producing nuclides in
the observed abundances.  Various problems  have been addressed in
recent studies that, for example, investigate spallation principally in
the circumstellar disk corona, consider production of meteoritic
isotopes such as $^{138}$La and $^{10}$Be in addition to well-known
isotopes such as $^{26}$Al and $^{41}$Ca, and examine the effects of
$^3$He-rich flares and self-shielding by chondrule rim
 material \citep{Clayton95, Lee98, Gounelle01, Goswami01}.  Spallogenic
isotope production by cosmic rays pervading star forming regions has
also been considered \citep{Clayton94}, but an early report of
$\gamma-$ray lines in the Orion region supporting this model has been
withdrawn \citep{Bloemen99}.

Some meteoritic abundance anomalies are more plausibly explained by
local spallation of solar nebular solids than by injection of stellar
nucleosynthetic material.  $^{21}$Ne is readily produced by proton
spallation reactions on common rock-forming elements like Mg and Si,
and is present with high abundance in certain grains of carbonaceous
chondrites \citep[e.g.][see review by Woolum \& Hohenberg 1993]
{Caffee87, Rao97, Caffee01}.  A spallation origin is supported by high
particle track densities in the crystalline grain.  This combination of
evidence implies that the grains were subject to high proton fluences
while free-floating in the nebula prior to meteoritic compaction.
Similarly, the spallogenic product $^{10}$Be has recently been reported
in the Allende meteorites consistent with a local irradiation model
\citep{McKeegan00}, although production in an anisotropic jet during a
supernova explosion is also possible \citep{Cameron01}.  The recent
discovery of $^{36}$Ar and other noble gases in silicate chondrites
also points to an enhanced solar wind prior to chondrule melting
\citep{Okazaki01}.

In addition to spallation, magnetic reconnection flares produce sudden
increases in the photon flux and (unobserved but plausible by analogies
with solar flares) shocks within the solar nebula environment.  These
may be responsible for the flash melting of chondrules \citep{Levy89,
Taylor92, Shu97, Shu01}.  Although local irradiation and shock models
often call for enhanced magnetic activity in the early Sun, the size,
location and geometry of the flaring magnetic fields is uncertain.
Field lines may be rooted in the proto-solar surface as in the
contemporary Sun \citep[e.g.][] {Feigelson91}, may link the proto-Sun
to the nebula at the corotation radius at $\sim 10$ R$_\odot$
\citep[e.g.][]{Shu97, Montmerle00, Birk00}, may erupt from an turbulent
magnetohydrodynamical (MHD) nebula \citep[e.g.][455ff]{Levy89,
Romanova98, Priest00}, or may be produced by interactions between the
solar wind and the nebula at planetary distances \citep{Cameron95}.
The magnetic geometry issue will be discussed here in \S
\ref{disk_sec}.

Astronomical studies of young solar analogs can provide valuable
insights into these issues.  For example, although injection freshly
synthesized material from a passing AGB star is an attractive
explanation for the origin of several short-lived nuclides
\citep{Busso99}, the chances are very small that an AGB star passes
sufficiently close to any given molecular cloud for this mechanism to
operate \citep{Kastner94}.  Triggering by supernova shocks does occur, but
may affect only a fraction of star forming sites \citep{Adams01}.

On the other hand, a large body of astronomical studies of young
solar-type stars supports the idea that high levels of magnetic flaring
are present in virtually all stars through all stages of pre-main
sequence evolution \citep[see review by][] {Feigelson99}.  This
provides an empirical basis to local spallation models for meteoritic
isotopic anomalies.  X-ray emission from pre-main sequence low-mass
stars is enhanced $10^1-10^4$ times above levels seen in the Sun today,
exhibiting high plasma temperatures ($T \sim 10^7-10^8$ K) and rapid
high-amplitude variability indicative of violent magnetic reconnection
events.  Nonthermal radio continuum flaring has been detected in a
smaller fraction of young stars \citep[see review by][]{Andre96}.  This
is demonstrably gyrosynchrotron emission from MeV electrons spiraling
in magnetic fields, direct evidence for particle acceleration to MeV
energies in pre-main sequence systems.  Extensive evidence (photometric
starspots, Doppler imaging, Zeeman line splitting, chromospheric
indicators, etc.) has also accrued indicating that T Tauri stars have
highly magnetized stellar surfaces.  The link between X-ray flaring in
T Tauri stars and meteoritic issues has been discussed by Feigelson
(1982), Feigelson et al.\ (1991), Shu et al.\ (1997), Feigelson \&
Montmerle (1999) and Priest \& Forbes (2000, 455ff).

Past studies of magnetic activity in analogs of the early Sun have
suffered practical limitations. Some individual $\simeq 1$ M$_\odot$
young stars have been studied in detail, such as V410 Tau or EK Dra
\citep[e.g.][]{Herbst89, Strassmeier98}, but it is unclear whether
these represent the average case or outliers with unusually rapid
rotation and magnetic activity.  X-ray emission was surveyed in young
stellar clusters in nearby star forming clouds with the $Einstein$ and
$ROSAT$ orbiting X-ray telescopes \citep[reviewed in][]{Feigelson99}.
But the resulting samples typically have $< 100$ stars across the
initial mass function with only a handful of solar analogs. Richer
young stellar clusters in more distant giant molecular clouds were
examined with $ROSAT$, but its optics could not resolve the crowded
fields and only the strongest flares could be detected.

We report here an observational study of the young solar analogs in the
richest unobscured young star cluster in the nearest giant molecular
cloud: the Orion Nebula Cluster, whose OB members ionize the Orion
Nebula (Messier 42).  The study is conducted with the recently-launched
{\it Chandra X-ray Observatory}.  $Chandra$'s superb mirrors provide
arcsecond spatial resolution, and the Advanced CCD Imaging
Spectrometer (ACIS) detector at its focal plane provides high
quantum efficiency with very low background noise and independent
spectroscopy of each star. The satellite has a high-ellipticity
 orbit that permits continuous observation over many hours.

Our principal goal here is to establish unbiased quantitative measures
of magnetic activity in young solar-like stars which can serve as
empirical input into local irradiation models for meteoritic isotopic
anomalies and shock models for chondrules formation.

\section{Observations and data reduction}

The Orion Nebula Cluster (ONC) is the richest young star cluster within
500 pc with $\simeq 2000$ members concentrated in a 1 pc ($8^\prime$)
radius sphere \citep{Odell01}.  The full initial mass function from a
45 M$_\odot$ O star to dozens of substellar brown dwarfs is present.
Most of the stars are optically visible (i.e.\ are not deeply embedded
in the Orion molecular cloud and are not heavily absorbed) with $V<20$
magnitudes.  These stars have well-characterized photometry and
spectroscopy from which bolometric luminosities and spectral types are
measured on the Hertzsprung-Russell diagram \citep{Hillenbrand97}.
Their locations on the HR diagram are then compared to theoretical
stellar interior models \citep{DAntona97} so that stellar ages and
masses can be inferred.  While most of the stars have ages $\leq 1$ My,
some have lower luminosities implying ages of $5-20$ My. An additional
population of deeply embedded stars lies behind the ONC around the OMC
1 cloud cores, but these will not concern us here.

The ONC was observed with the ACIS-I imaging array on board {\it
Chandra} twice during the inaugural year of the satellite, on 12 Oct
1999 and 1 Apr 2000, for $\simeq 12$ hours on each occasion.  The
satellite is described by \citet{Weisskopf96} and details of the ACIS
instrument are available at \url{http://www.astro.psu.edu/xray/acis}.
\citet{Garmire00} gives a report of the first observation, and a full
description of the data reduction and sources found in both
observations is given by \citet[][hereafter F01]{Feigelson01}. We
summarize these procedures briefly here.

The satellite telemetry data were corrected for charge transfer
inefficiency, cleaned of cosmic ray afterglows, hot columns,
undesirable grades and energies.  The two observations were merged, and
astrometrically aligned to the Hipparcos frame using 2MASS sources to
$\pm 0.1^{\prime\prime}$ accuracy.  Source detection based on  a
wavelet transform algorithm located 1075 sources with a sensitivity
limit of $9$ counts ($L_x \simeq 1 \times 10^{28}$ erg s$^{-1}$ for a
typical unabsorbed spectrum) near the field center.  Source photons
were extracted from the 95\% or 99\% encircled energy circle except
where crowding is present.  Background is negligible in most cases and
is not subtracted here. Effective exposure times include effects of
telemetry dropouts, telescope vignetting, and point spread function
wings.  The total exposure time near the field center is 83 ks.
Least-squares spectral fitting to the pulse height distributions were
performed assuming a Raymond-Smith plasma with soft-energy absorption,
all assuming cosmic elemental abundances.  A two-temperature plasma
model was adopted when needed for satisfactory spectral fits.  Most of
the observed emission is attributable to optically thin plasma emission
with $T \simeq 10-30$ MK (MK = $10^6$ $^\circ$K), though some plasma
with $T \geq 100$ MK can be present.  The ACIS-I detector is
insensitive to plasma cooler than $T \leq 3 $ MK.

The merged ACIS-I field is shown in Figure \ref{image_fig} in the
$0.5-8$ keV total energy band.  This is the richest field of X-ray
sources ever obtained in the history of X-ray astronomy.  92\% of the
X-ray sources are unambiguously identified with optical/infrared ONC
cluster members with angular offsets $<1$\arcsec\/ in the inner
10\arcmin\/ where the telescope point spread function is optimum.  For
the X-ray and optically bright stars under discussion here, there is
little doubt that the stellar identifications are correct with one
caveat: most of the stars are probably binary or multiple systems at
angular scales smaller than $Chandra$'s arcsecond resolution
\citep{Mathieu94}.  In the case of visual binaries (those with `a' or
`b' in their names in Table \ref{analogs_tab}), we associate the X-ray
source with the brighter member of the system (see footnotes of Table 2
in F01 for individual cases).

\section{Selection and X-ray properties of ONC solar analogs}

\begin{figure} 
\plotone{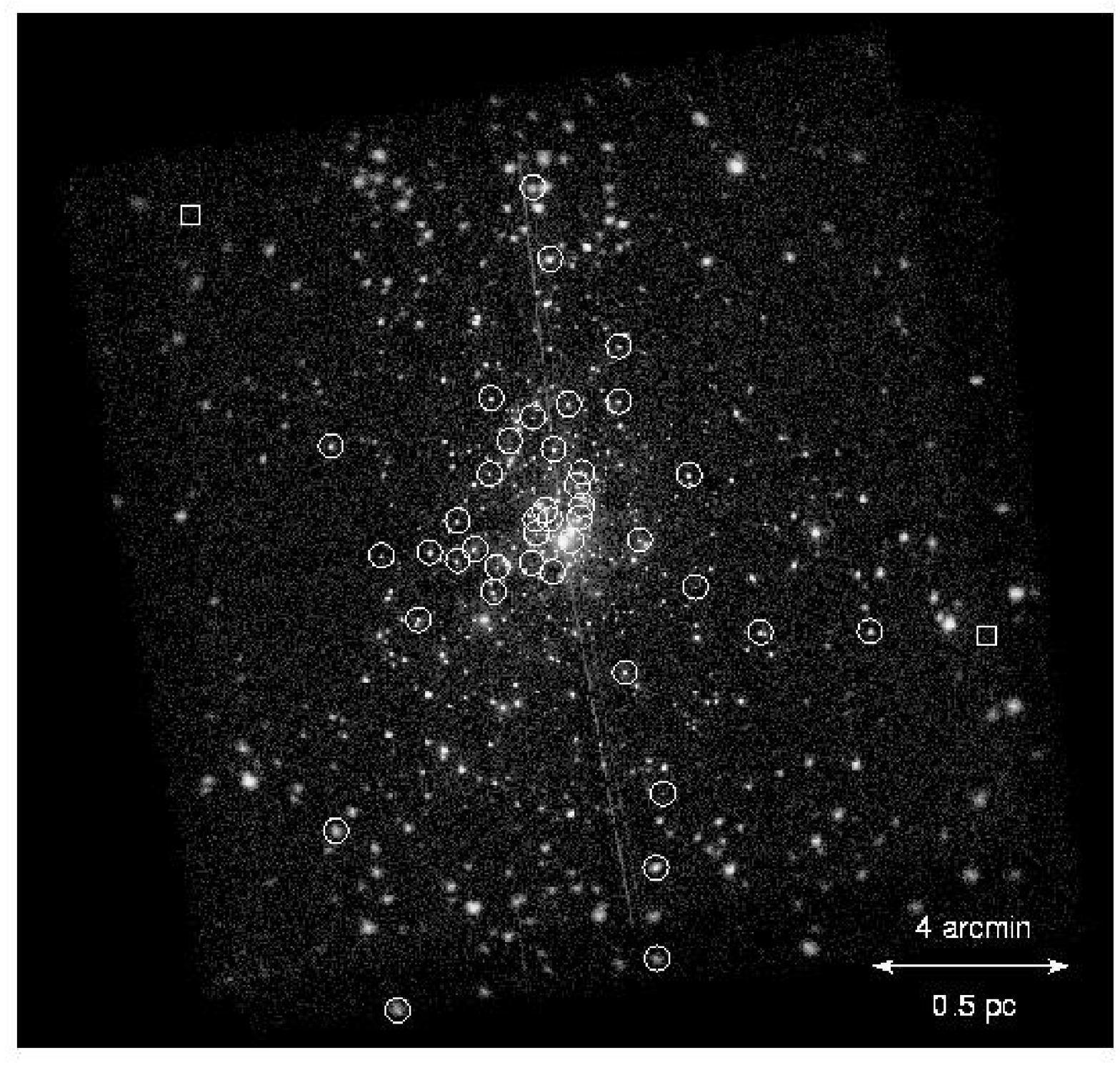} 
\vspace{0.5in}
\caption{The Chandra ACIS-I image of the Orion Nebula Cluster with 43
$0.7 \leq M \leq 1.4$ M$_\odot$ stars marked.  The two boxes indicate
the two undetected stars.  The field of view is about $17\arcmin\times
17\arcmin$, the telescope resolution is about 1$\arcsec$~ near the
field center, and the brightest source at the field center is the O6
star $\theta^1$C Ori. The image intensity here is proportional to the
log of the photon density of the image, shown here at reduced
resolution. See F01 for a full resolution map of the field.
\label{image_fig}}

\end{figure}

In this study of analogs of the pre-main sequence Sun, we select for
study all well-characterized ONC stars with $V<20$
\citep{Hillenbrand97} with inferred masses in the range $0.7 \leq M <
1.4$ M$_\odot$.  Mass is chosen as the critical discriminant because it
is the strongest predictor of X-ray emission in pre-main sequence stars
\citep{Feigelson01b}. These 43 stars are listed in Table
\ref{analogs_tab} and marked in Figure \ref{image_fig}.  All but two,
shown as squares in Figure \ref{image_fig}, are associated with ACIS
X-ray sources.  This is a unique ensemble of pre-main sequence with
masses very close to 1 M$_\odot$ that have uniform distances, excellent
optical/infrared characterization, and nearly complete X-ray
detections from a survey with nearly uniform sensitivity.

The sample is complete for unobscured portions of the ONC, and includes
a few of moderately absorbed stars.  It suffers one selection bias:
roughly $10-20$ ONC stars seen from $JHK$ infrared images lie in our
mass range but are sufficiently deeply embedded that their visual
magnitudes $V > 20$ \citep{Hillenbrand00}.  Unless some surprising link
between large-scale gaseous environment and X-ray emission is present,
the omission of these embedded stars from the sample studied here
should have no effect on our results other than the removal of $\simeq
20$ random objects from the full cluster sample of solar-mass stars. We
therefore consider the stars listed in Table \ref{analogs_tab} to be an
unbiased sample of solar-mass pre-main sequence stars.

Table \ref{analogs_tab} lists the sample stars with relevant
optical/infrared and X-ray properties.  Most entries are extracted from
Tables 2 and 3 of F01.  Columns $1- 2$ give the ACIS source number from
F01 and its IAU designation.  Column 3 gives the stellar counterpart
from \citet{Jones88}, and columns $4-6$ give the luminosity, visual
absorption and estimated age from \citet{Hillenbrand97}. Note that we
have truncated ages below $\log t < 5.5$ yr due to the uncertainties in
theoretical modeling of the initial conditions and effects of accretion
\citep[e.g.][]{Wuchterl01}.  Column 7 summarizes the evidence
concerning a circumstellar disk: a {\bf +} symbol indicates a
near-infrared photometric excess $\Delta(I-K) > 0.3$ and/or association
with a Herbig-Haro outflow, far-infrared source or imaged proplyd; a
{\bf $-$} symbol indicates $\Delta(I-K) < 0.3$ and no association of
these types; and $\ldots$ indicates insufficient information for
classification.  Column 8 gives the extracted ACIS counts in the
$0.5-8$ keV band which are used to evaluate the X-ray properties.  For
the two undetected stars, upper limits to the extracted counts and
X-ray luminosities were obtained following the procedure in \S 2.11 of
F01.  Columns $9-10$ give the count rates in the first and second ACIS
exposures respectively.  Column 11 gives the variability class defined
in F01:  Constant (Const), Long-Term Variable (LT Var) indicating
significantly different count rates between the two exposures, Possible
flare (Pos fl) and Flare indicating variability within one or both
exposures.  Columns $11-13$ give fitted spectral parameters:
interstellar column density along the line of sight in H atoms
cm$^{-2}$, plasma energy in keV (with two energies listed if a
one-temperature model provided a poor fit to the spectrum), and an
indicator that Spectral Features (S.F.) are present, possibly due to
non-solar elemental abundances in the heated plasma.

Column 14 gives the X-ray luminosity $\log L_x$ in the total band
derived by integrating the spectral model over the $0.5-8$ keV band and
multiplying by $4 \pi d^2$ with $d = 450$ pc.  This is called $L_t$ in
F01 and is corrected for differences in source spectrum, off-axis point
spread function and telescope vignetting losses but is not corrected
for line-of-sight absorption. F01 also give soft band, hard band and
absorption-corrected total band luminosities for each source.  These
values typically lie within $\pm 0.4$ of $\log L_x$ and show the same
behavior in scatter plots such as Figure \ref{evol_fig}.  The $\log
L_x$ values are averaged over both $\simeq$12-hour ACIS exposures
which, as seen in Figure \ref{lightcurves_fig}, masks substantial
variability of the sources.  Considerable scatter in $\log L_x$ values
is thus expected due to the limited temporal coverage of each star's
flaring behavior, which is unavoidable in any study of high-amplitude
stellar flaring.

\section{X-ray evolution and flaring in ONC solar analogs
\label{results_sec}}

We first evaluate the time-averaged X-ray luminosities of ONC solar
analogs to provide an overview of the level and longevity of enhanced
magnetic activity during the pre-main sequence phase. Then the
short-term variations are examined to quantify the intensity and
frequency of magnetic reconnection flares.

\subsection{Long-term evolution of pre-main sequence X-rays
\label{evol_subsec}}

\begin{figure}
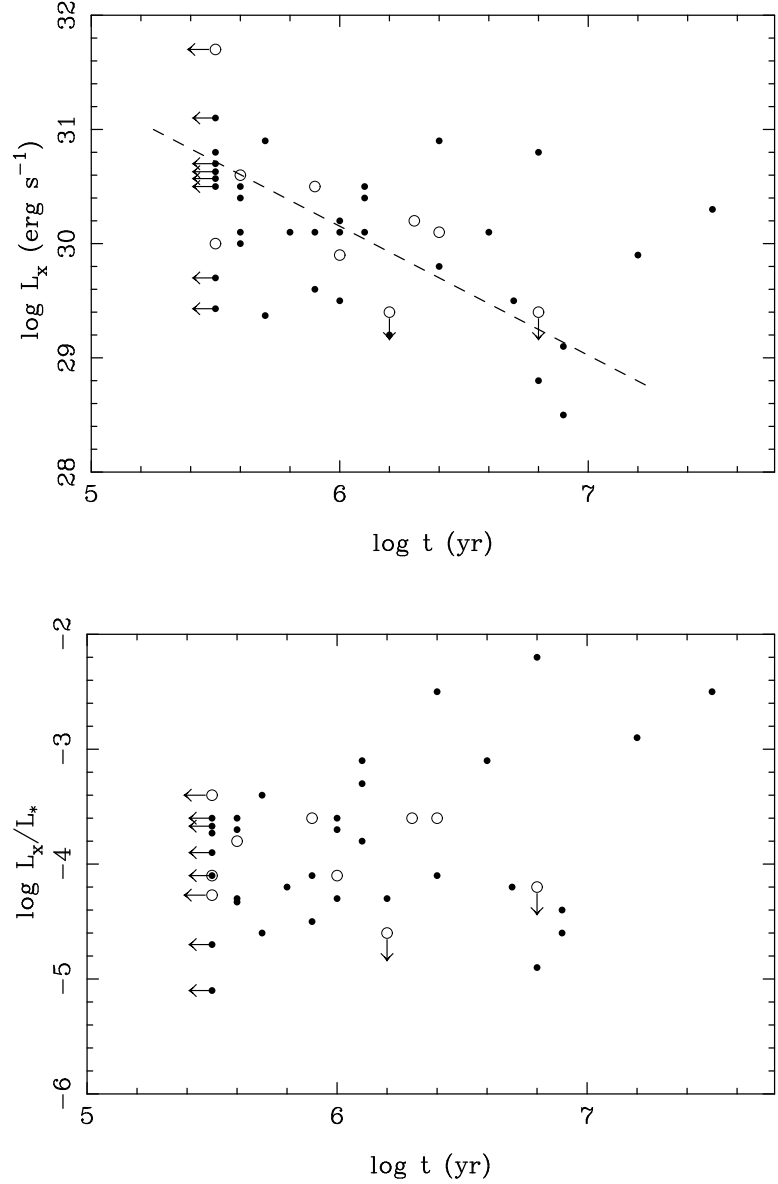

\centering
  \begin{minipage}[t]{1.0\textwidth}
  \centering
  \includegraphics[height=4.0in,angle=-90.]{f2a.eps}
  \end{minipage} \\ [0.3in]
  \begin{minipage}[t]{1.0\textwidth}
  \centering
  \includegraphics[height=4.0in,angle=-90.]{f2b.eps}
  \end{minipage}
\caption{(a) Plot of the time-averaged X-ray luminosity in the $0.5-8$ keV band
against the estimated stellar age for ONC solar analogs.  (b) Plot of the ratio 
between X-ray and bolometric luminosities against age. Filled circles
represent stars with evidence for circumstellar disks (or with no information on 
disks), and open circles represent stars without disks. 
\label{evol_fig}}
\end{figure}

The distribution of X-ray emission for solar-mass ONC stars as a
function of disks and stellar age is shown in Figure \ref{evol_fig}.
The mean\footnote{The mean, population standard deviation, and tests
for correlation are evaluated using survival analysis techniques which
take into account the two nondetections (left censored data points).
See Chapter 10 of \citet{Babu96} for methodological details and code
ASURV at \url{http://www.astro.psu.edu/statcodes} for implementation.}
X-ray luminosity is  $<\log L_x> = 30.1 \pm 0.5$ erg s$^{-1}$ for the
full sample  and $30.3 \pm 0.5$ erg s$^{-1}$ for the stars $\leq 1$ My
in age.

But Figure \ref{evol_fig}a suggests that the X-ray luminosity
decreases as the star descends the Hayashi pre-main sequence tracks.
An anticorrelation between $L_x$ and age is present at a significance
level of $P \simeq 0.003$, as evaluated by the generalized Spearman's
$\rho$ and Kendall's $\tau$ nonparametric tests for correlation$^1$.  A
linear regression fit\footnote{Here we use the ordinary least squares
(OLS) bisector line which is the most reliable of OLS lines that treats
the two variables symmetrically.  Regression coefficient errors ($1
\sigma$) are estimated both analytically and with bootstrap
simulation.  Upper limits are considered to be detections for this
calculation.  See Chapter 7.2 of \citet{Babu96} for methodological
details and code SLOPES at \url{http://www.astro.psu.edu/statcodes} for
implementation.}, shown as the dashed line, gives

$$ \log L_x = 30.2 (\pm 0.8) ~-~ 1.1 (\pm 0.1) \log (t/10^6~{\rm yr})~~
{\rm erg ~s}^{-1}.  $$

This finding suggests that magnetic activity decreases as the star ages
from $\log t \simeq 5$ to 7 yr, but the correlation is based on only a
few stars.  This result was absent in early $Einstein$ $Observatory$
results \citep{Walter91} and was tentatively reported in one $ROSAT$ T
Tauri sample \citep{Feigelson93}, while another $ROSAT$ study showed a
rise in $L_x$ up to $\log t \simeq 6.5$ yr followed by a decline
\citep{Neuhauser95}.  

We note from Figure \ref{evol_fig}a that the presence or absence of a
circumstellar disk (filled or open circles) has no discernable effect
on the level of X-ray emission.  This result has been found in many
X-ray studies of pre-main sequence stars \citep[][F01]{Feigelson99}, 
though diskless stars were found to be brighter in X-rays than
classical T Tauri stars in a $ROSAT$ study of the Taurus-Auriga 
clouds \citep{Stelzer01}. 

Figure \ref{evol_fig}b shows the evolution of the ratio $\log
L_x/L_\star$, which is very closely related to the X-ray surface flux
$\log L_x/(4 \pi R_\star^2)$. Here the mean value is $\log L_x/L_\star
= -3.9 \pm 0.5$ ($1 \sigma$) for the whole sample and there is no
statistically significant age dependence ($P \simeq 0.3$). However, as
noted by \citet{Garmire00} in a preliminary analysis of this diagram,
the magnetic activity of older stars with $\log t > 6.0$ yr has a
higher dispersion ($\pm 0.8$) than younger stars with $\log t \leq 6.0$
yr ($\pm 0.4$).  The causes of this are unknown: neither the presence
of a disk nor surface rotation accounts for this effect
\citep{Feigelson01b}.  Perhaps stars follow two tracks as they age, one
group with roughly constant $\log L_x$ and rising $\log L_x/L_\star$
and another group with declining $\log L_x$ and roughly constant $\log
L_x/L_\star$.  In this hypothetical case, we could not ascertain
whether or not the early Sun participated in the general decline of
$\log L_x$ as it aged.  Note that a plot of the mean $log L_x/L_\star$
and $\log t$ for all (mostly sub-solar mass) stars in several young
stellar clusters based on $ROSAT$ data shows a factor of $\sim 10$ rise
in $\log L_x/L_\star$ with age from $5 < \log t < 8$ \citep{Kastner97},
suggesting that only some stars participate in the falling $\log L_x$
we find in Figure \ref{evol_fig}a. For the purposes of this study, we
conclude only that X-ray emission in solar analogs is typically $\log
L_x \simeq 30.0 - 30.5$ erg s$^{-1}$ during the first 1-2 Myr.

\subsection{Short-term temporal behavior of pre-main sequence X-rays}

The Variability Class listed in Table \ref{analogs_tab} is a
qualitative indicator of the degree and nature of the variations.
Twenty-eight of the 41 detected stars (68\%) of the sources show
intraday variability (classes `Flare' and `Possible flare').  Of the
remaining 13 sources (classes `Constant' and `Long-Term Variable',
three show high-amplitude (factor $>3$) differences between the two
observations suggesting that a long-duration flare is present during
the high flux observation.  Most of the remaining sources are weak with
reduced sensitivity to rapid flares.  We thus find that while 2/3 of
the sources directly show variations on timescales of hours, the true
fraction may be substantially higher.  It is difficult to establish
this fraction quantitatively due to the wide range of source count
rates.

\begin{figure}
\centering
  \begin{minipage}[t]{1.0\textwidth}
  \centering
\includegraphics[scale=0.30]{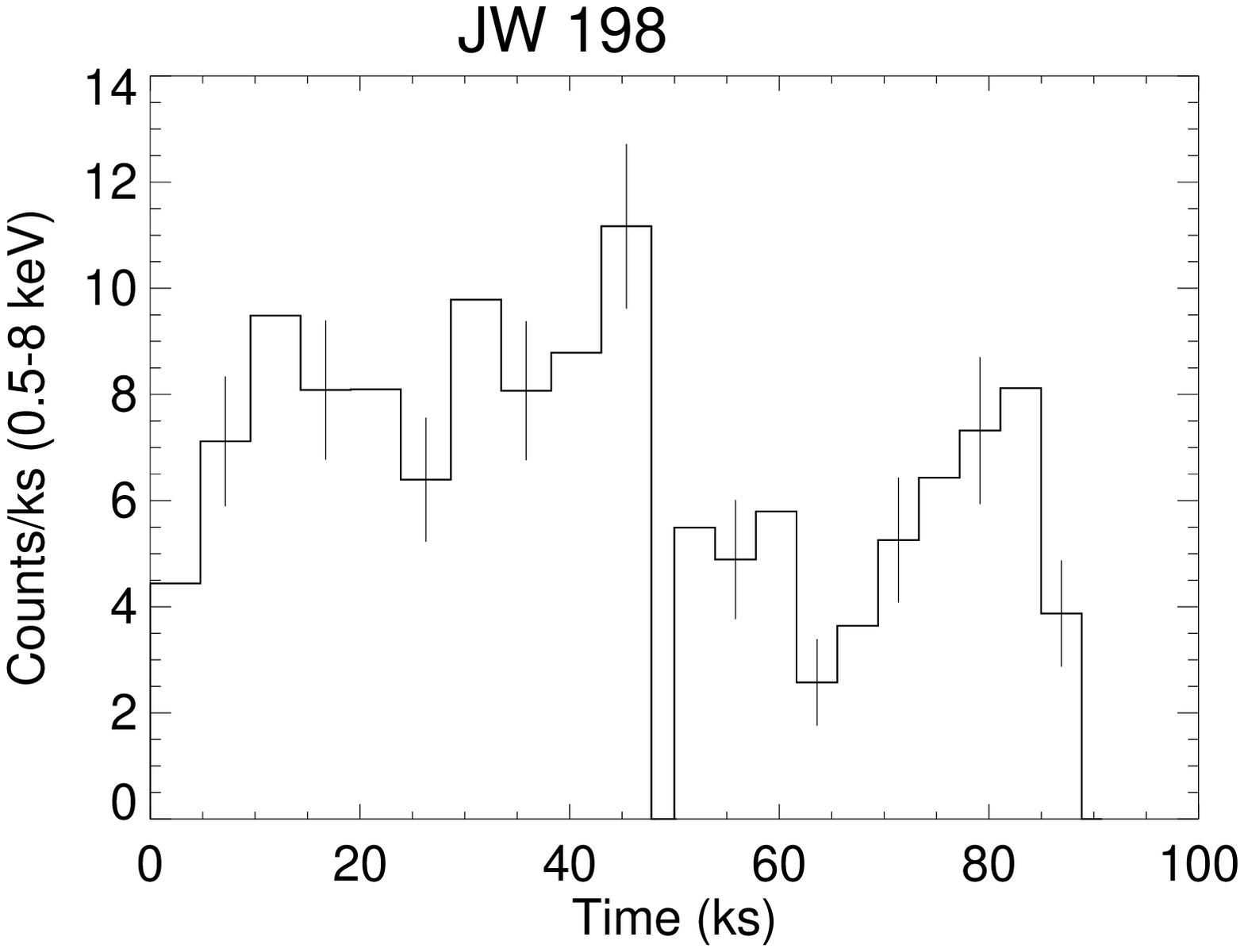}
\includegraphics[scale=0.30]{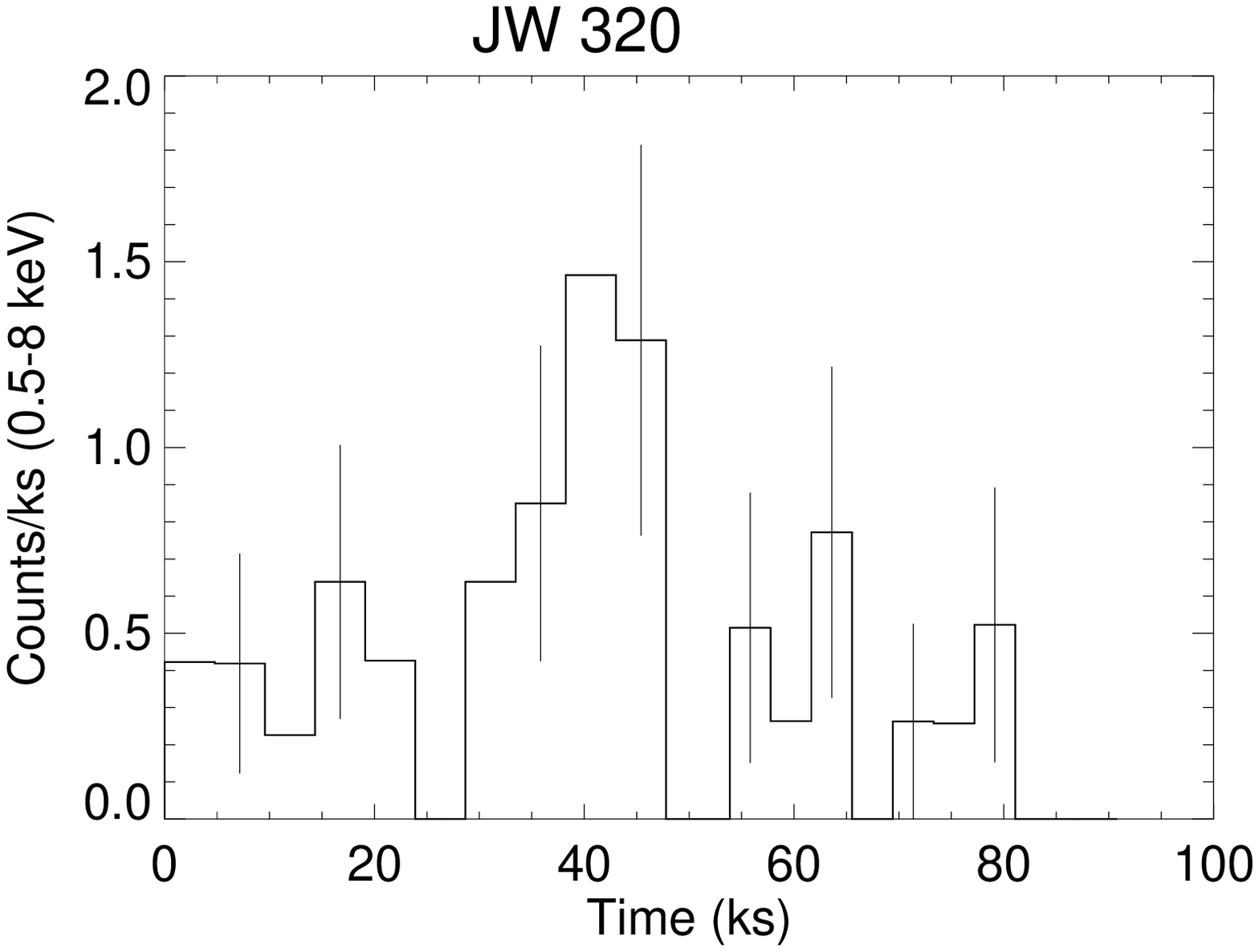}
\includegraphics[scale=0.30]{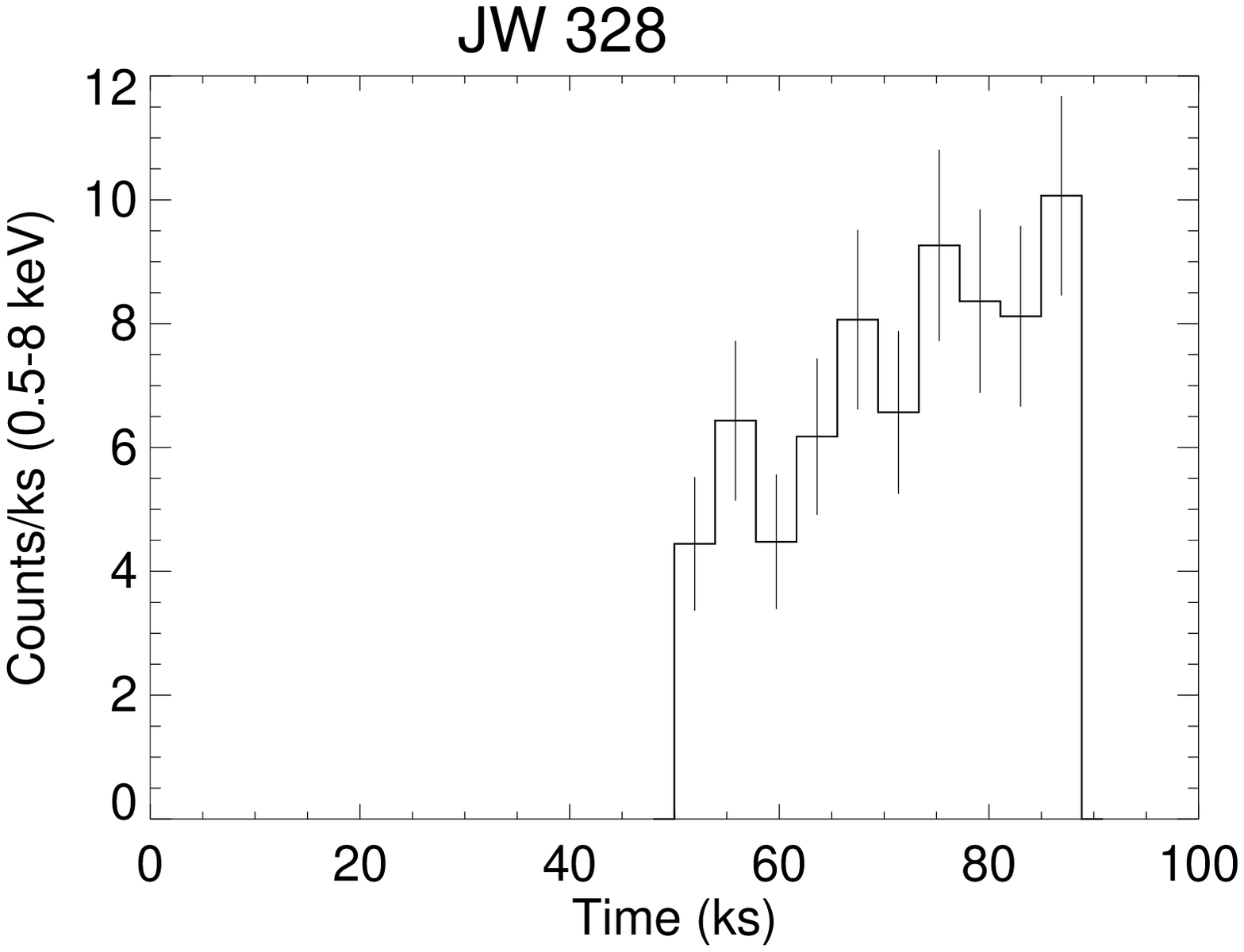}
  \end{minipage} \\ [0.1in]
  \begin{minipage}[t]{1.0\textwidth}
  \centering
\includegraphics[scale=0.30]{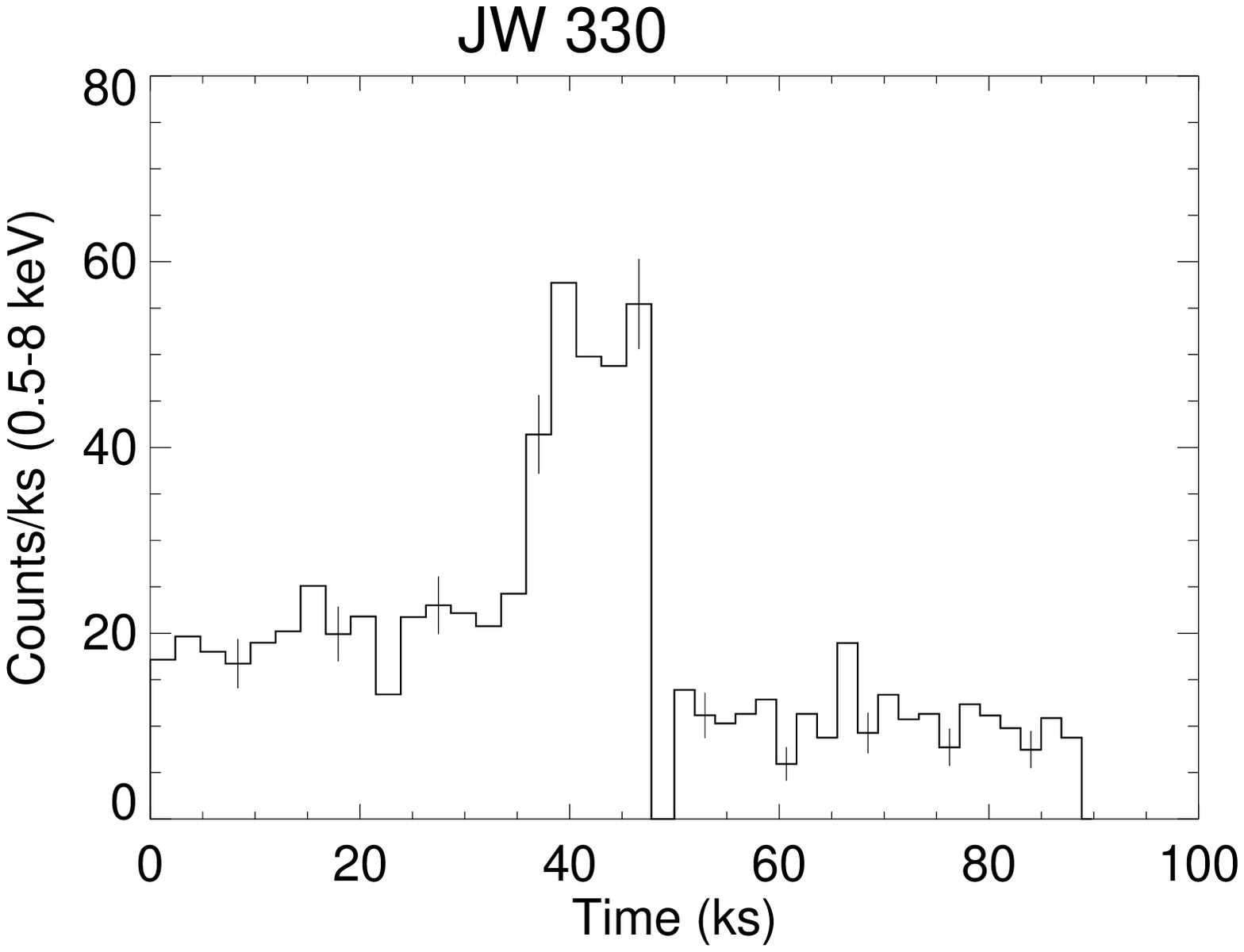}
\includegraphics[scale=0.30]{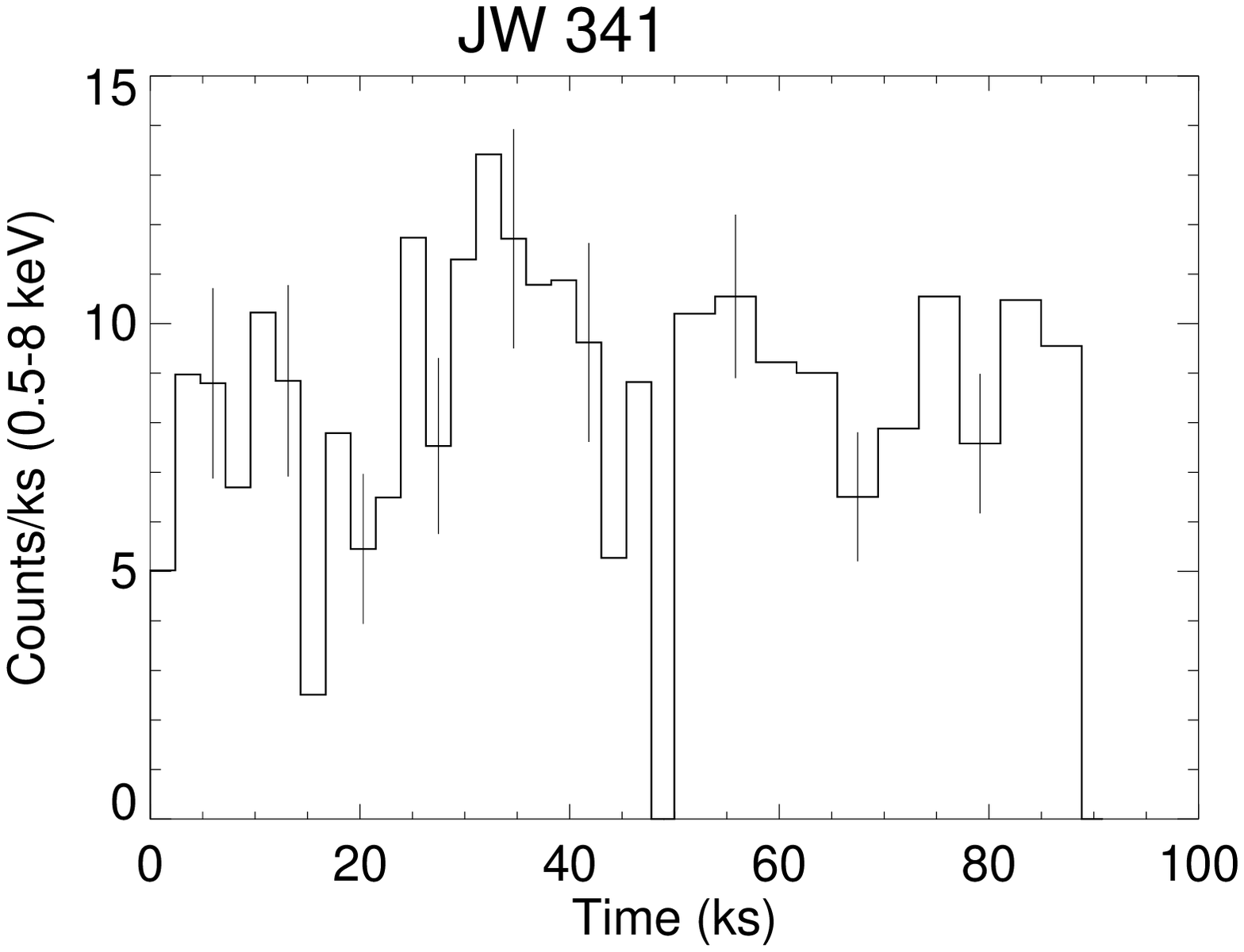}
\includegraphics[scale=0.30]{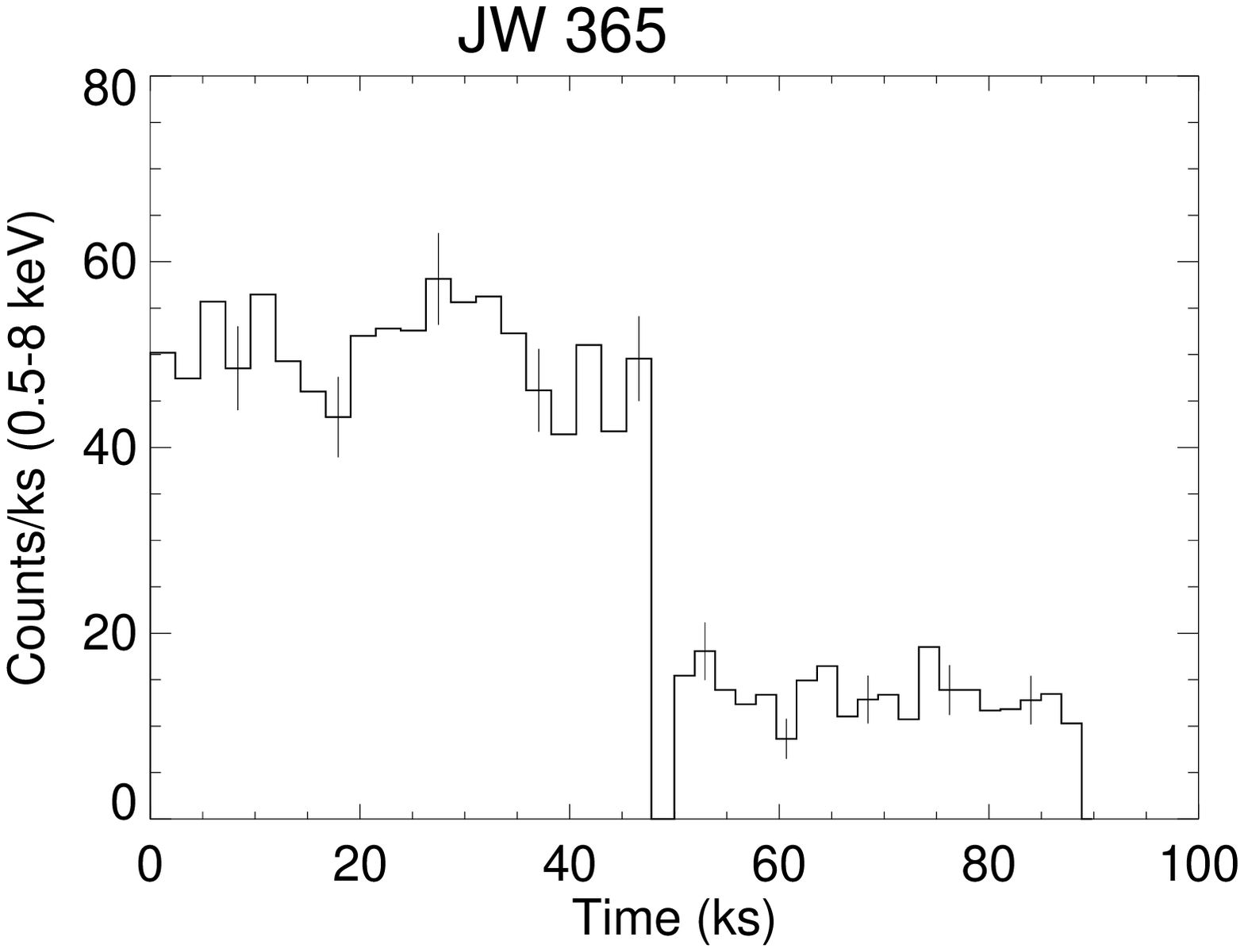}
  \end{minipage} \\ [0.1in]
  \begin{minipage}[t]{1.0\textwidth}
  \centering
\includegraphics[scale=0.30]{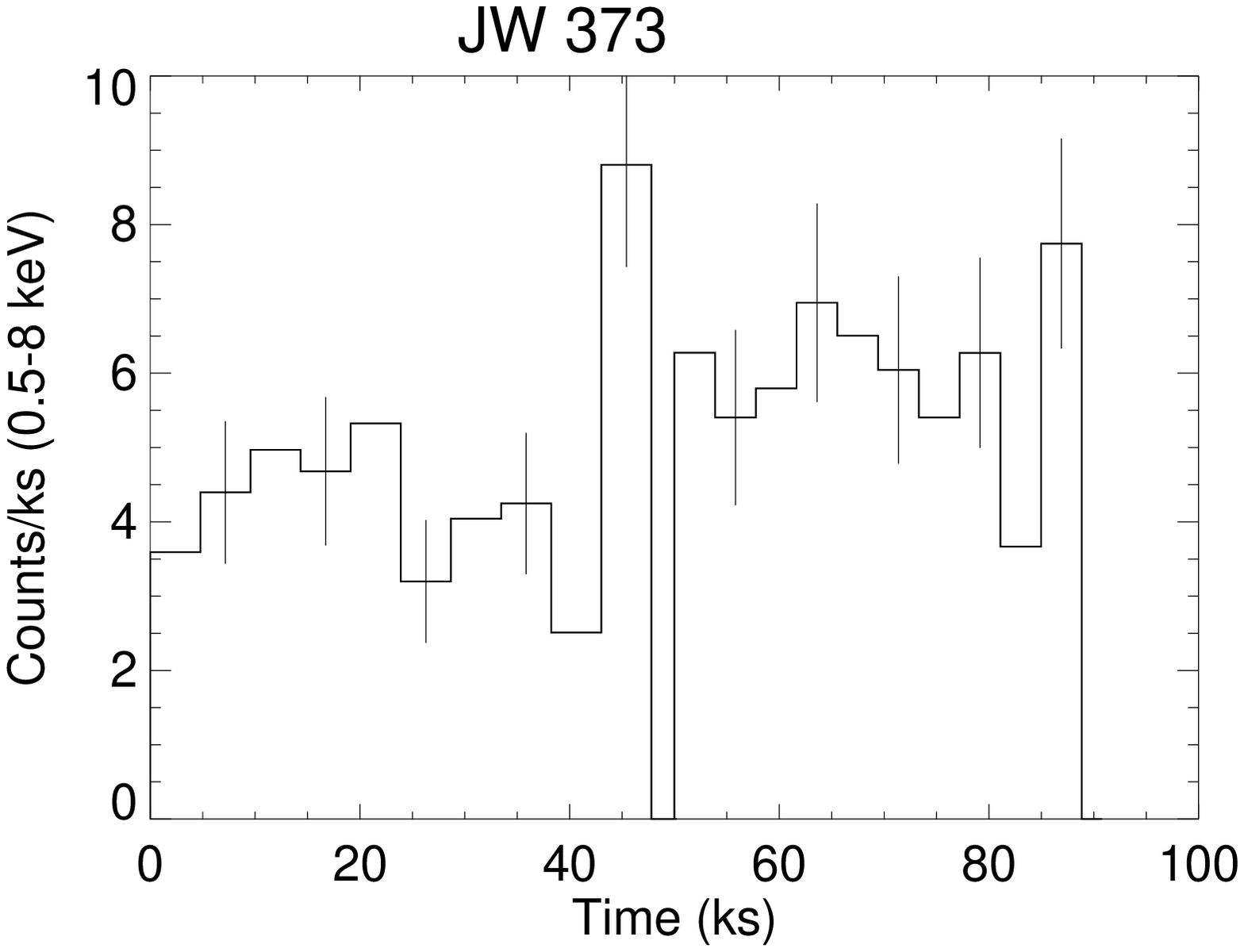}
\includegraphics[scale=0.30]{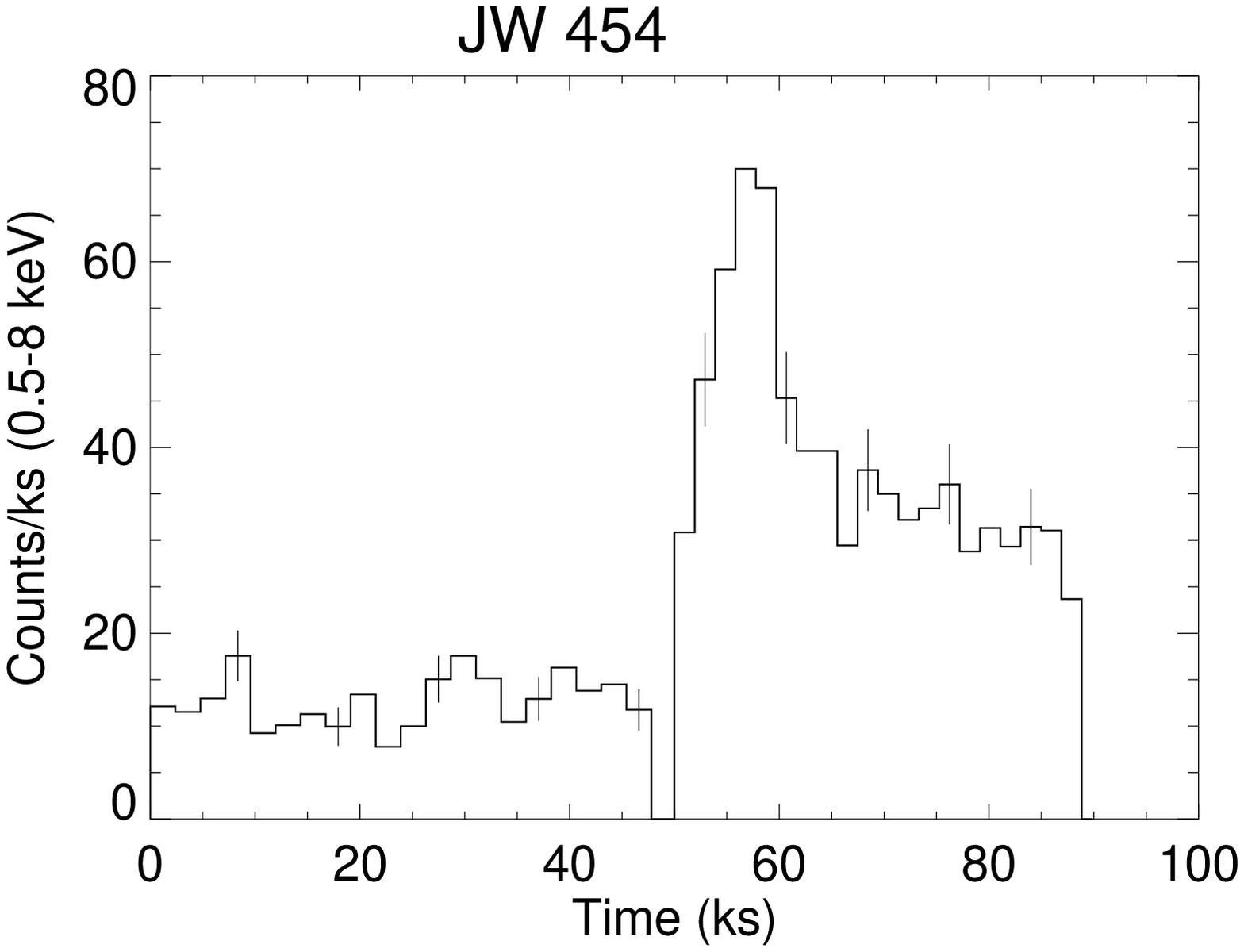}
\includegraphics[scale=0.30]{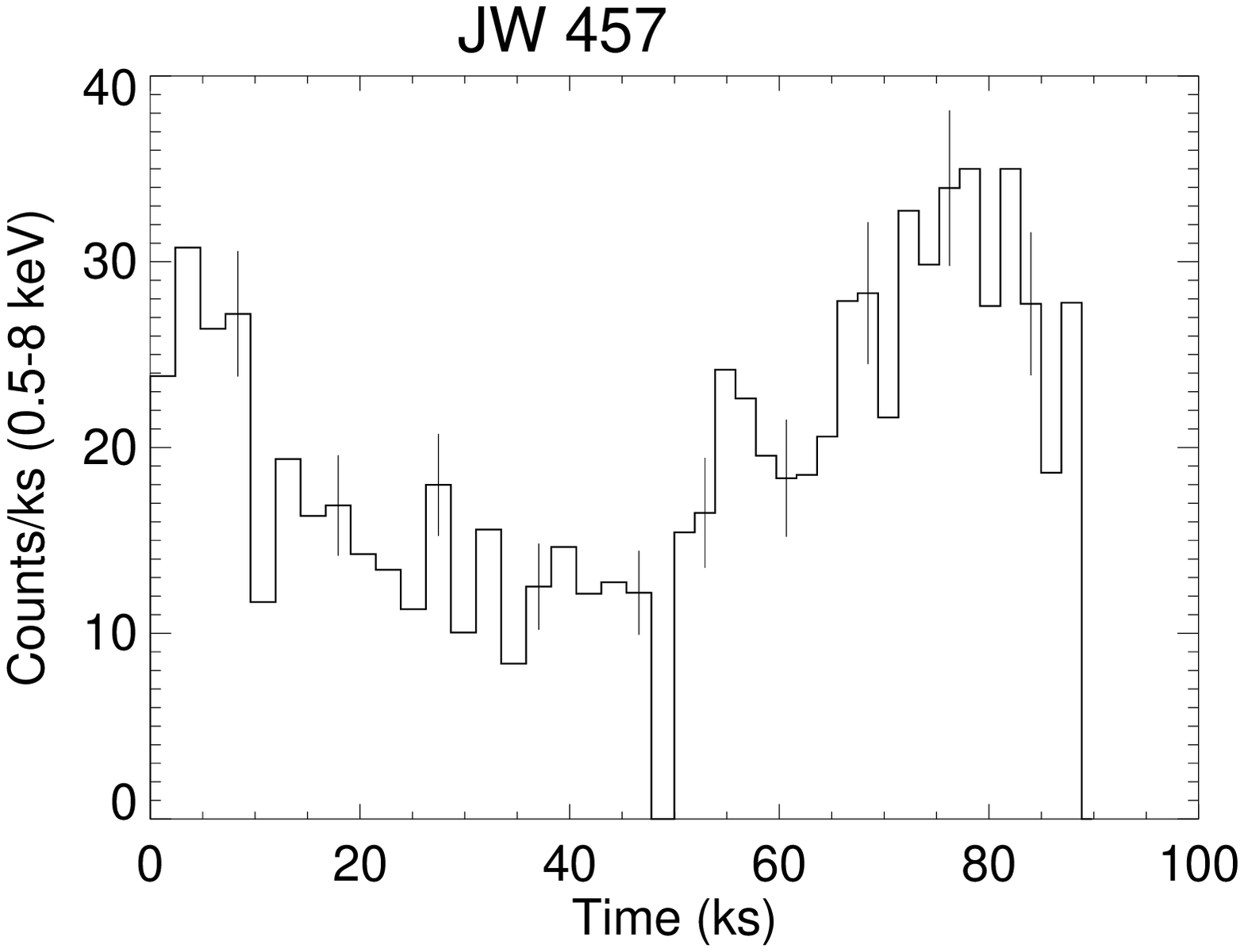}
  \end{minipage} \\ [0.1in]
  \begin{minipage}[t]{1.0\textwidth}
  \centering
\includegraphics[scale=0.30]{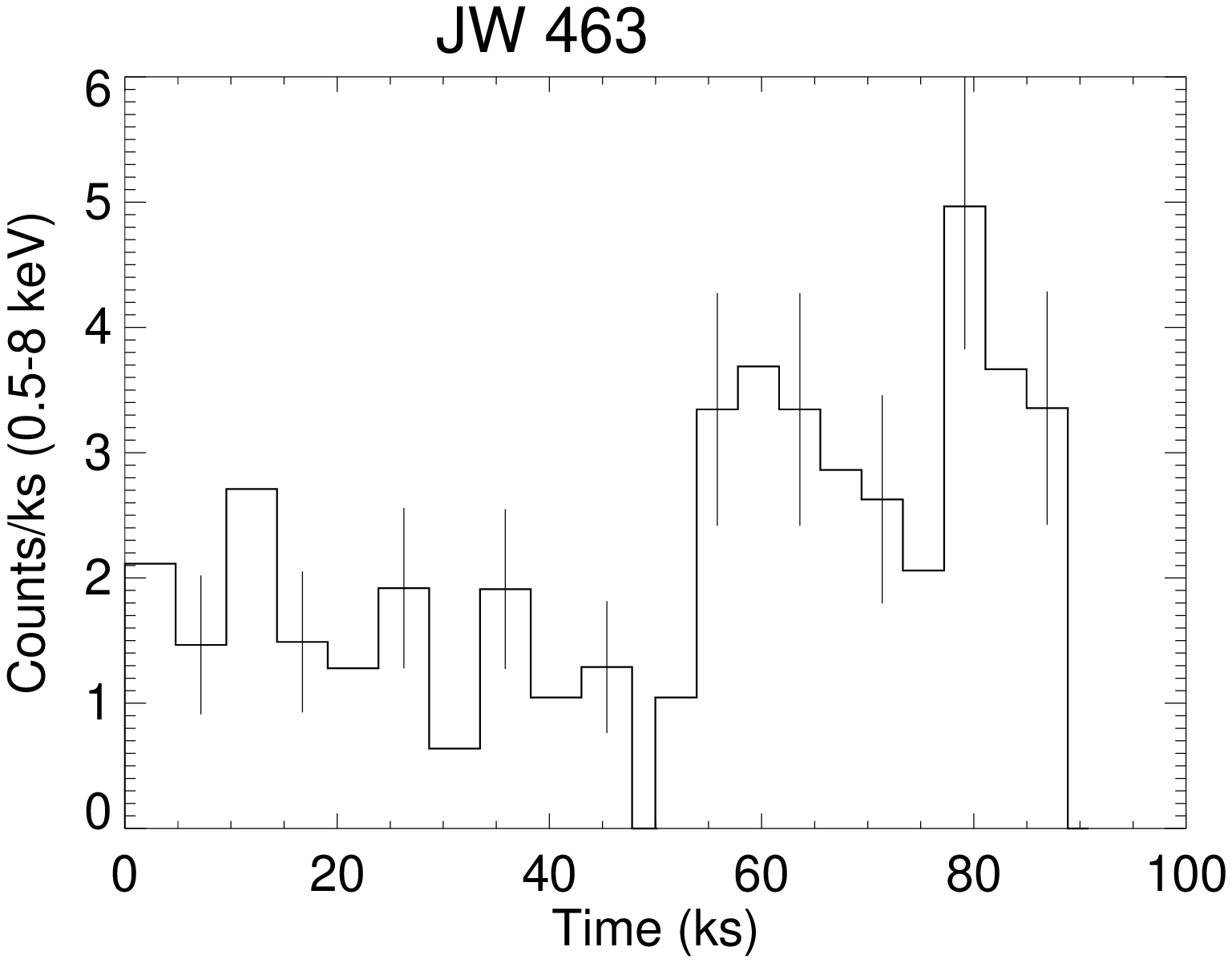}
\includegraphics[scale=0.30]{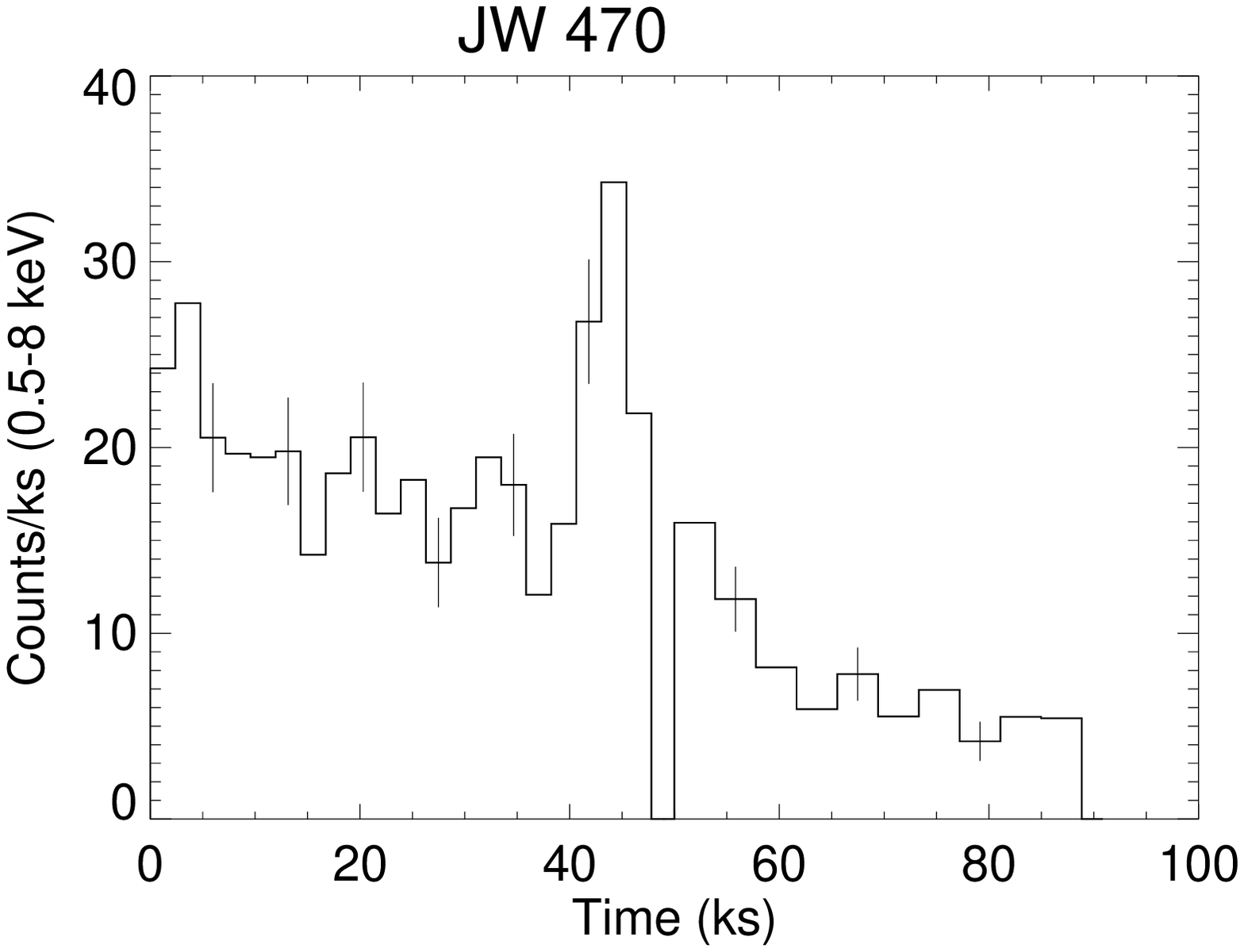}
\includegraphics[scale=0.30]{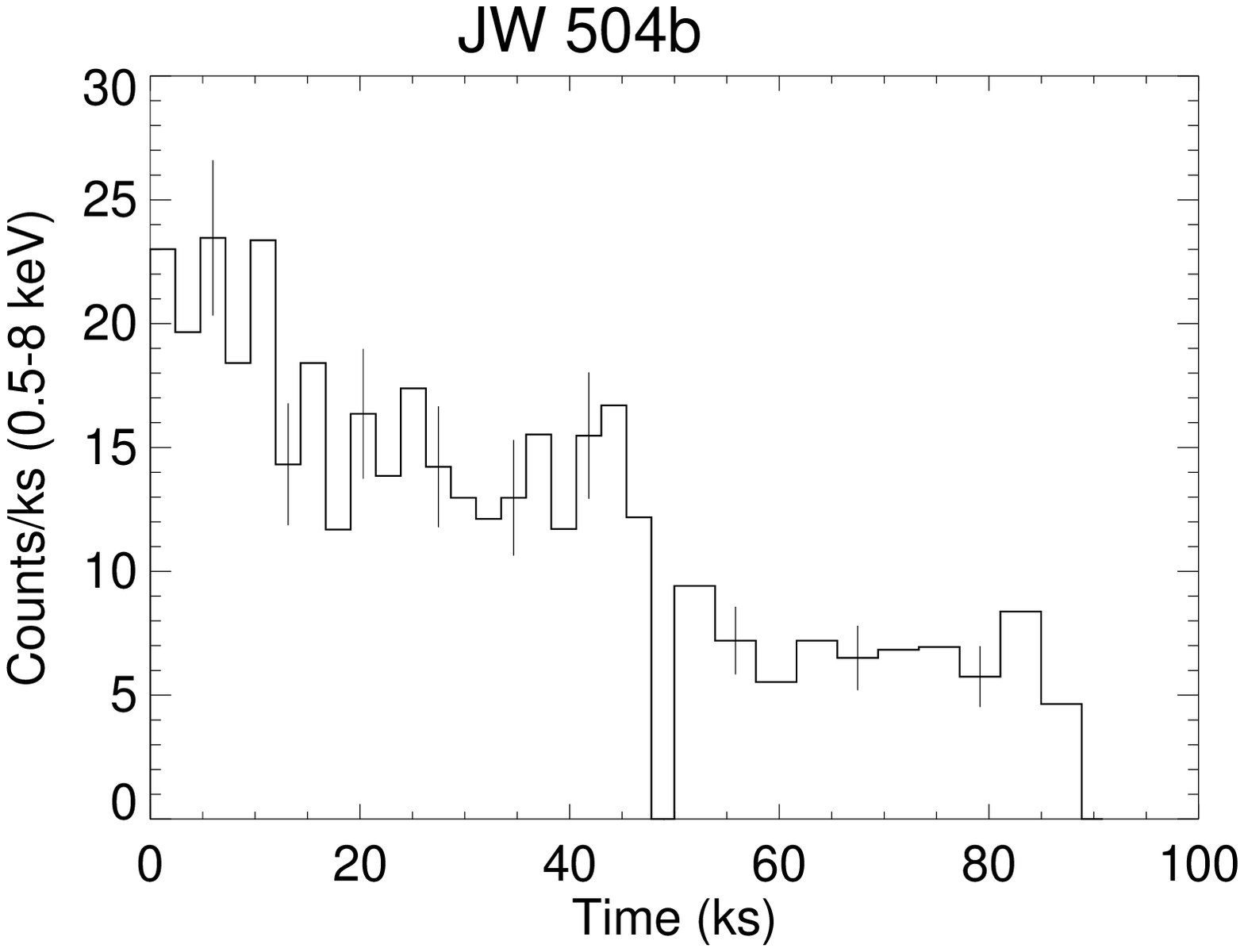}
  \end{minipage} \\ [0.1in]
  \begin{minipage}[t]{1.0\textwidth}
  \centering
\includegraphics[scale=0.30]{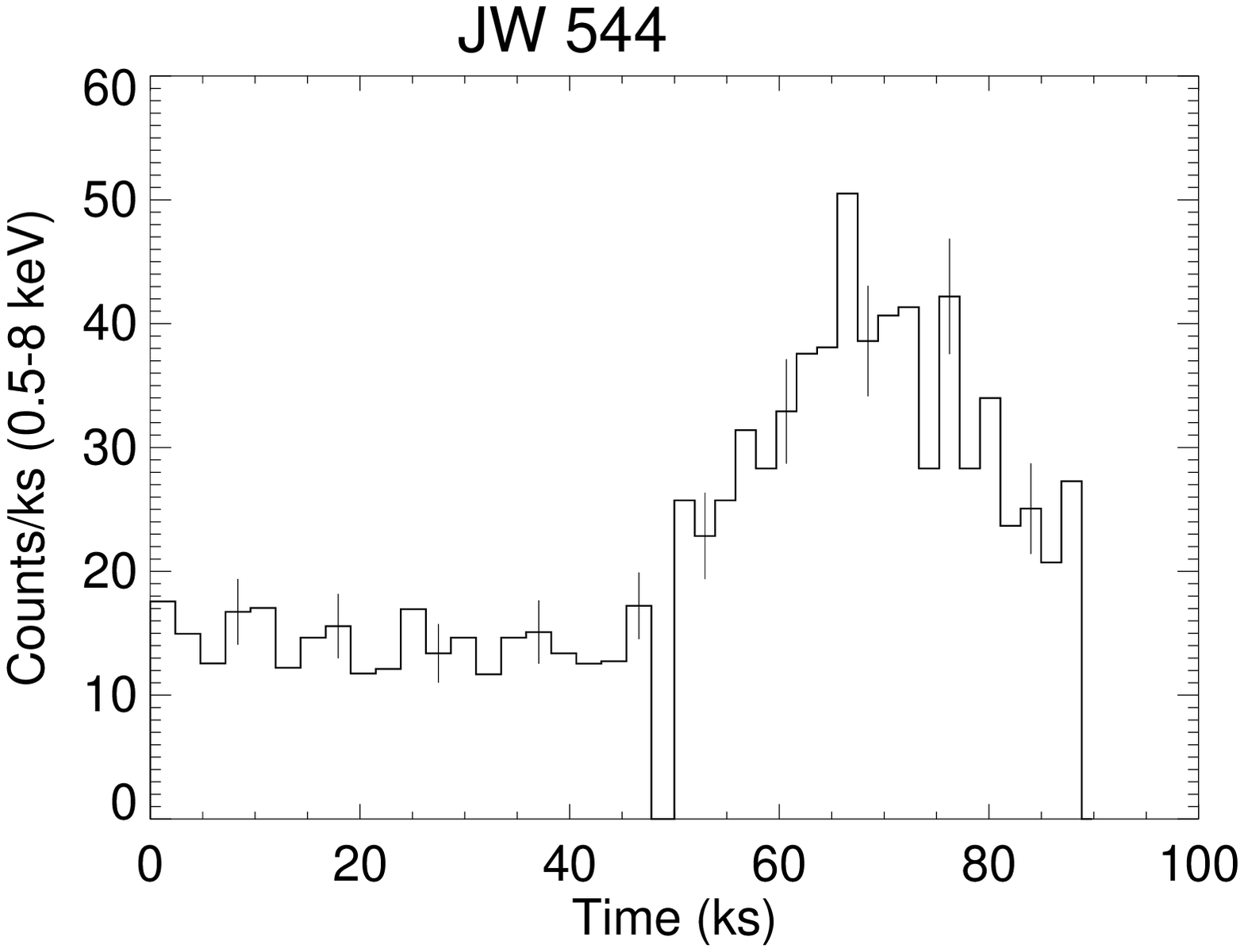}
\includegraphics[scale=0.30]{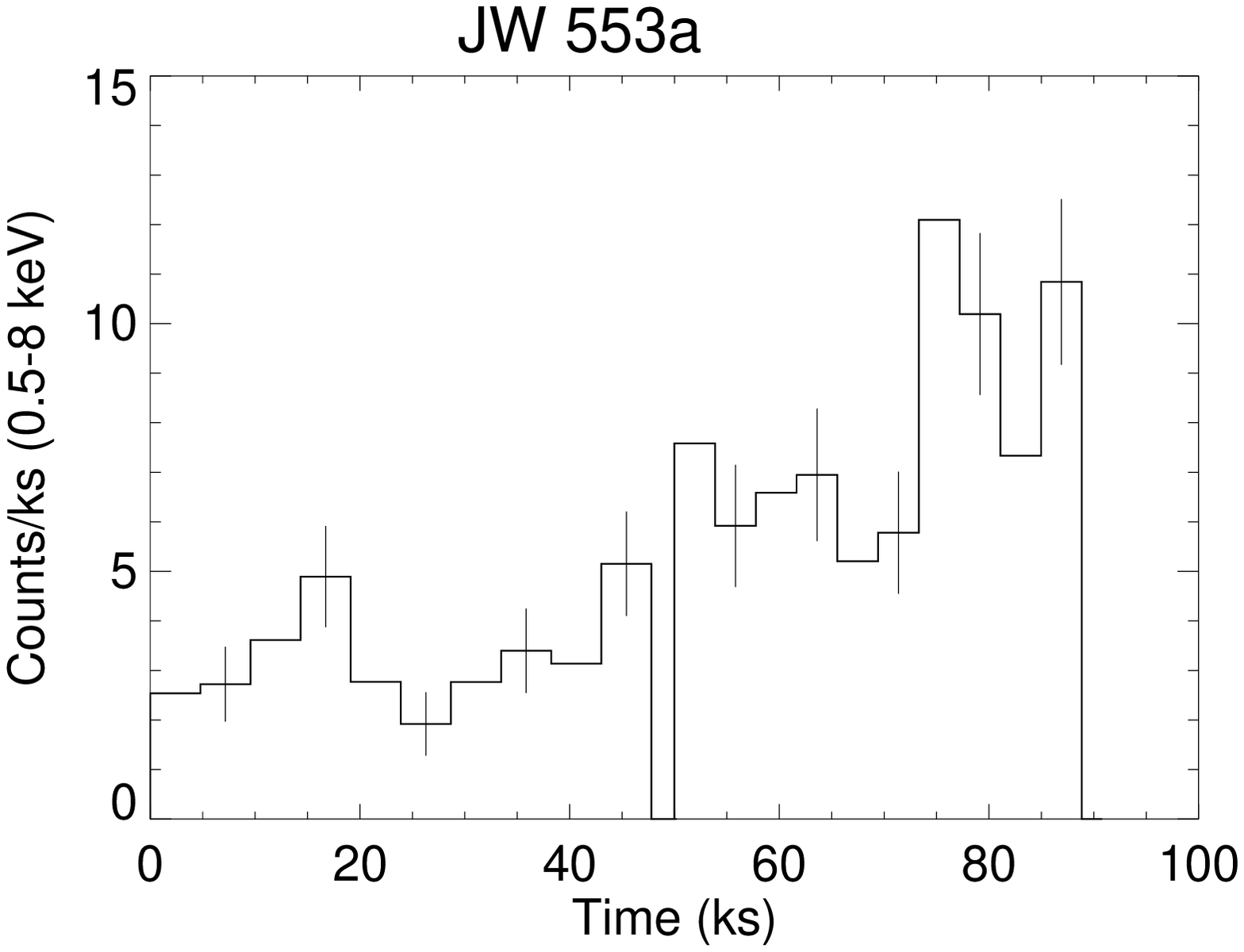}
\includegraphics[scale=0.30]{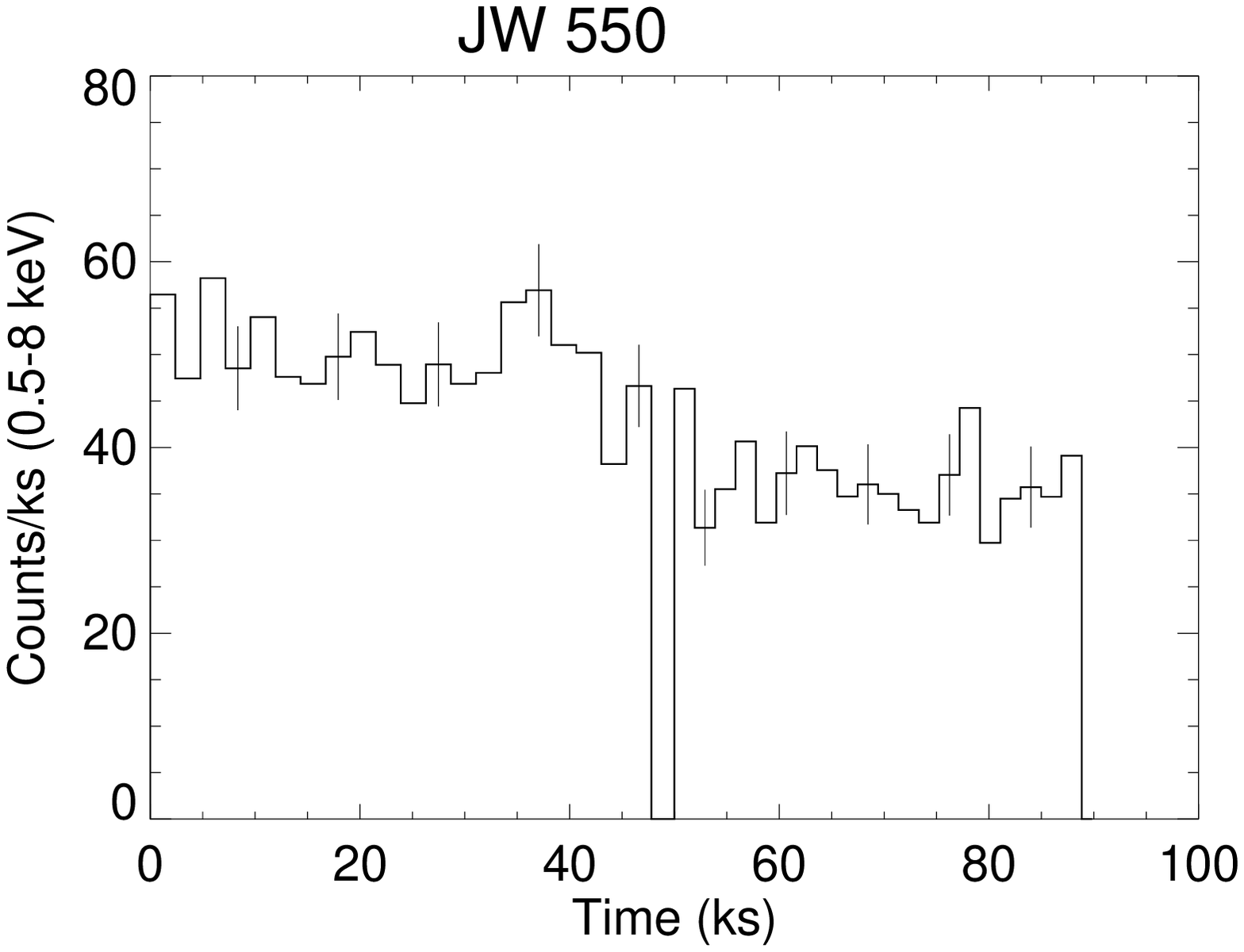}
\vspace{-0.2in}
\caption{X-ray lightcurves of solar analog stars in the Orion Nebula
Cluster exhibiting intraday variability (Variability Class `Possible flare' and 
`Flare').  For convenience, the two observations obtained on 12 Oct 1999 and 1 
Apr 2000 are concatenated such that the second observation starts at Time = 50 
ks. JW 328 and JW 847 were outside the field of view during one observation.
Binwidths are set in each panel to give about 20 counts per bin; typical 
$\sqrt{N}$ uncertainties are shown. \label{lightcurves_fig}}
   \end{minipage}
\end{figure}

Figure \ref{lightcurves_fig} shows the X-ray lightcurves for the 28
stars exhibiting intraday variability and for two stars (JW 553a and JW
907) exhibiting high-amplitude long-term variability.  The diversity of
temporal behaviors is evident from these 30 lightcurves.  Some show
fast high-amplitude flares with relatively brief durations of $1-4$
hours superposed on a 'quiescent' level (e.g.  JW 454, 698 and 810),
others show slow high-amplitude changes on timescales $\geq 12$ hours
likely to be long-duration flares (e.g.  JW 328, 504b, 567, 596, 738
and 826).  A few cases show two events during the same 12-hour interval
(e.g. JW 457, 463, 769, 826).  Some of the less dramatics variations,
as well as those of stars not shown, may be attributed to rotational
modulation of static coronal structures.

While one cannot reliably distinguish individual flares from quiescent
levels in all sources, we estimate from Figure \ref{lightcurves_fig}
that 30 flares with $29.0 \leq \log L_x(peak) \leq 31.5$ erg s$^{-1}$
were present in the ONC solar analogs during our observations.  Their
power is weighted towards the more rather than less luminous flares: 20
of these 30 events have $\log L_x(peak) \geq 30$ erg s$^{-1}$.  Total
energies in the X-ray band, $E_x = L_x t_f$ where $t_f$ is the duration
of the flares, range from $10^{32}$ to $>10^{36}$ erg.  Flaring is
present in both older and younger pre-main sequence stars, those with
and without  circumstellar disks, and those with slow as well as rapid
surface rotation.  It is a ubiquitous phenomenon.  

\section{ONC flares, solar flares and inferred proton fluences
\label{inferred_sec}}

The results from our Chandra ACIS observations of the ONC solar
analogs  most salient to the early Sun and solar nebula irradiation
issues are the average luminosity, peak flare luminosities, and flare
frequcies in comparison to levels seen on the Sun today.

We find that all solar-mass stars with ages $\leq 1$ My, averaged over
both temporal variations and stars with ages $\leq 1$ My, show an
average X-ray luminosity of $<\log L_x> = 30.3$ erg s$^{-1}$ in the
$0.5-8$ keV band.  For comparison, in the same band the contemporary
Sun emits typically $\log L_x \simeq 25.3$ erg s$^{-1}$ with $T \simeq
1$ MK during its quiet phase, $\log L_x \simeq 27.0$ erg s$^{-1}$
during maximum phase with $T \simeq 2-5$ MK, and around $\log L_x
\simeq 28.0$ erg s$^{-1}$ with $T \simeq 2-20$ MK during a powerful (2
Nov 1992) flare \citep{Peres00}.  The ONC solar analogs thus
collectively show a factor of $\sim 10^4$ elevation in X-ray emission
over the Sun averaged over the solar cycle, and $10^{2.5}$ elevation
over the most luminous solar flares.

The prevalence of rapid X-ray variations and high plasma temperatures
of $10 \leq T \leq 100$ MK indicate that pre-main sequence X-rays
generally arise from flare rather than coronal phenomena, even when
individual flare events are not clearly seen in the
lightcurves\footnote{Most of the observed X-ray emission in ONC stars
is also too hot to arise from accretion onto the stellar surface from
the disk or environs \citep{Lamzin99}.  See, however, \citet{Kastner01}
for a soft-spectrum T Tauri star where accretion may dominate the X-ray
spectrum.}.  For comparison, the most X-ray luminous flares of recent
solar cycles had $\log L_x (peak) \simeq 28.5 \pm 0.5$ erg s$^{-1}$ in
the $Chandra$ spectral band (see Appendix).  Taking $10^{30}$ erg
s$^{-1}$ for the typical flare in our ONC sample, we find that ONC
solar analogs exhibits flares $10^{1.5}$ times more X-ray luminous than
these extreme solar events.  The ONC flares are roughly $10^4$ times
more powerful than solar flares which occur with a daily frequency.

Our ONC dataset is collectively equivalent to 86 0.5-day observations
of a single solar-mass pre-main sequence star spread over a long period
of time.  The observed ONC flare frequency is thus about 30 flares over
43 days or 1 flare every 1.4 days or $1 \times 10^5$ s.  The maximal
solar flares with $\log L_x (peak) \simeq 28.5$ erg s$^{-1}$ occur
roughly once every year or $3 \times 10^7$ \citep[][see
Appendix]{Sammis00}.  The ONC flares we detect are thus occur roughly
$10^{2.5}$ times frequently than the most powerful solar flares.

We can now estimate the time-averaged enhancement of X-ray flare
luminosity of ONC solar analogs compared to the strongest contemporary
solar long duration events: $10^{1.5}$ elevation in peak luminosity
times $10^{2.5}$ more frequent occurrence gives a $10^4$ enhancement in
total flare X-ray emission.  This is similar to the $10^4$ factors we
find comparing the ONC stars with the time-averaged X-ray luminosity
and daily solar flares.

For application to computations of nuclear spallation reactions, we
must consider that solar proton fluences scale non-linearly with solar
X-ray luminosity in the sense that most of the protons are produced by
the most powerful flares.  This is usually expressed in terms of the
frequency distribution of events as a function of energy $E$: the
distribution of proton events, $dN/dE \propto E^{-1.15}$, is
significantly flatter than the distribution of X-ray events, $dN/dE
\propto E^{-1.6}-E^{-1.8}$ \citep{Hudson91, Crosby93}.  If this
difference extends over the $10^2$ extrapolation in luminosity between
solar and ONC flares, one infers an additional $10^1$ increase in
proton fluences in the early Sun from the observed giant
flares\footnote{A final multiplicative factor associated with the
proton contribution of flares fainter than the $\log L_x \simeq 29.0$
erg s$^{-1}$ flare sensitivity limit of the Chandra observations can be
included.  But this may be relatively small because the frequency
distribution of solar proton events is heavily weighted towards the
strongest flares so the weaker flares, though more numerous, contribute
relatively little to the total particle fluence.}.  We note that there
is no astrophysical basis for this extrapolation; it is possible that
proton fluence saturates at some lower level, as appears to occur in
the contemporary Sun \citep[][see Appendix]{Reedy96}.  However, the
existence of efficient acceleration of relativistic particles by
pre-main sequence flares is observationally established by their radio
gyrosynchrotron emission (\S \ref{protons}).

The enhancement in proton fluences of $<1$ My ONC solar analogs is thus
inferred to be about $10^5$ above that produced by the most powerful
flares from the Sun today: $10^{1.5}$ from the increased X-ray
luminosity, $10^{2.5}$ from the increased occurrence frequency, and
$10^1$ from the increased proton flux compared to X-ray flux.  As the
contemporary proton fluence at 1 A.U. from the flare site averaged over
a solar cycle is measured to be $f_p \simeq 200$ protons cm$^{-2}$
s$^{-1}$ with energies $E>10$ MeV \citep[][see Appendix]{Reedy96}, the
inferred proton fluence at 1 A.U. for solar-mass stellar systems during
their first million years is $f_p \sim 10^7$ protons cm$^{-2}$
s$^{-1}$.  Note that, as the duration of a single powerful solar proton
event is typically several hours and flares with $\log L_x(peak) > 28$
erg s$^{-1}$ (below our detection limit) probably occurred several
times a day, it is likely that the pre-main sequence Sun was producing
high fluences of relativistic protons nearly continuously from
overlapping magnetic flare events.

It is useful for some theoretical calculations to estimate the
luminosity of energetic protons rather than the fluence at 1 A.U.  The
conversion is not simple as proton ejection and propagation is guided
by the magnetic fields and is not isotropic.  If we roughly assume
protons are ejected in an outward-facing hemisphere and that their
average energy is $E_{20} = 20$ MeV,  then the luminosity of $E>10$ MeV
protons inferred from the ONC flaring is $L_p \sim 2 \pi d^2 f_p E_{20}
= 5 \times 10^{29}~(d/{\rm 1 A.U.})^2$ erg s$^{-1}$.  The ratio
$L_p/L_x$ we infer for pre-main sequence flares is of order $10^{-1}$,
consistent with available knowledge of the energetics of powerful solar
flares.

We also note that the directly observed value  $<\log L_x> \simeq 30.3$
erg s$^{-1}$ should be useful for computations of X-ray ionization and
astrochemistry of the solar nebula and other circumstellar material
around solar-mass pre-main sequence stars \citep[e.g.][]{Igea99,
Aikawa99, Glassgold00, Glassgold01}.

\section{From Orion X-ray flares to meteoritic isotopic anomalies}

In this section, we discuss in some detail the degree of confidence in
the line of reasoning linking the observations of ONC X-ray flares and
the measurements of short-lived isotopic abundance anomalies in
carbonaceous chondrites.  We find that some links in the logical chain
are strong while others suffer considerable uncertainty.

\subsection{Do Orion results apply to the early Sun?
\label{Ori_Sun_sec}}

Our results show that the great majority of stars in our sample
exhibit  very high flaring levels during their first Myr.  A critical
element of our study's design is the completeness of the sample of
pre-main sequence solar mass stars within well-defined observational
criteria.  The only selection bias within the ONC is the preference for
unobscured stars in the evacuated region of the Messier 42 H{\sc II}
region over obscured stars lying in the undisturbed molecular cloud
behind the H{\sc II} region.  We have not thought of any mechanism
whereby this bias can affect magnetic activity.

While the X-ray luminosities of young stars have dependencies on mass
and age which we have accounted for, there is no known dependence on
their star forming environments.   Although quantitative studies are
lacking due to insufficiency of optical characterizations, $Chandra$
studies of  T Tauri stars in environments as sparse as the L1551 cloud
in Taurus-Auriga (Bally et al., in preparation) and NGC 1333 region in
Perseus \citep{Getman02} and as rich as the Rosette OB association
(Townsley et al., in preparation) have similar X-ray luminosity ranges,
plasma temperatures and variability properties.  With $\sim 2000$
stars, the ONC is roughly midway in log-mass between the richest star
clusters and the sparse groups forming in low mass nearby molecular
clouds.  When equal logarithmic intervals of mass are considered, a
given star has roughly equal probability of having formed in large and
small clusters \citep{Elmegreen97}.  The flaring activity of ONC
solar-mass stars can thus be applied to any solar mass star with
considerable confidence.

The universality of enhanced flaring stands in contrast to competing
models for external seeding of radioactive isotopes which require
special conditions.  For example, seeding by the passage of an AGB star
during its dredge-up phases when freshly synthesized atoms will
permeate its atmosphere \citep{Wasserburg94, Busso99} is likely to
apply in only $1:10^6$  star forming sites \citep{Kastner94}.

\subsection{Do X-ray flares produce energetic protons? \label{protons}}

In solar flares, the magnetic reconnection processes that heat plasma
to X-ray temperatures can also accelerate and release charged particles
with MeV energies (see Appendix).  Particle acceleration is evidenced
by variable nonthermal and circularly polarized radio continuum
emission, gamma-ray emission lines from spallogenic nuclear excitation,
and sudden enhancements of energetic particles propagating through the
interplanetary medium.  The processes are complex and not fully
understood. The heating of plasma to X-ray temperatures and the
acceleration of particles to MeV energies are sometimes nearly
simultaneous and cospatial, but are often decoupled.  For example, a
multiwavelength study of the magnetically active T Tauri star V773 Tau
showed several-fold decay of a radio flare simultaneous with a constant
high level of X-ray emission over several hours  \citep{Feigelson94}.
In the Sun, radio bursts associated with flares can occur on timescales
of one second to many hours and span $\geq 4$ orders of magnitude in
frequency \citep{Bastian98}.   Only during the brief impulsive phases
of some flares are radio, gamma-ray and hard X-ray emission nearly
simultaneous and clearly originate from the same reconnection event.

However, despite the complexity of solar flare physics and the lack of
a simple relationship between X-ray flares and solar energetic
particles, the link is statistically strong.  The NOAA Space
Environment Center uses the integrated solar luminosity above $1 \times
10^{26}$ erg s$^{-1}$ (M5 level) in the $1.5-12$ keV band from the
$GOES$ X-Ray Sensor to forecast the proton fluence that arrives at 1
A.U.  hours to days later \citep{Bornmann96}.

A strong statistical link also exists between X-ray emission and radio
gyrosynchrotron from energetic electrons in the solar-stellar arena.
The relationship $\log L_x \simeq 19.0 + 0.73 \log L_r$, where $L_r$ is
expressed in erg s$^{-1}$ Hz$^{-1}$, is found over 10 orders of
magnitude in $L_r$, extending from solar microflares to the most
magnetically active late-type stars \citep{Benz94}.  Radio
gyrosynchrotron emission has been detected including several dozen
weak-lined T Tauri stars at levels around $10^{25}$ erg s$^{-1}$ Hz$^{-
1}$ at centimeter wavelengths, and in a few cases high fractions of
circular polarization conclusively demonstrate that the emission is
gyrosynchrotron from trans-relativistic electrons accelerated in strong
magnetic fields \citep[see reviews by][]{Andre96, Feigelson99}.
Gyrosynchrotron emission is generally undetected from the earlier
evolutionary phases, probably due to free-free absorption by ionized
material associated with accretion and outflows.  But variable circular
polarized emission has been seen in three protostars, two of which also
show X-ray flares \citep{Phillips93, Feigelson98, Getman02}.  Young
stars with the most powerful flares, including weak-lined T Tauri
stars, generally show radio emission $1-2$ orders of magnitude above
the level predicted by the Benz-G\"udel relation.  This is attributed
to saturation of X-ray flares but not radio flares near the stellar
surface \citep{Gudel97}.

As it is difficult to imagine a magnetic reconnection process that
accelerates electrons but not protons, these radio findings provide
strong evidence that MeV protons are copiously produced in pre-main
sequence flares.

\subsection{Do the protons hit the disk? \label{disk_sec}}

There is considerable uncertainty about the spatial relationships
between disk solids and energetic flare particles in young stellar
system.  While a vast literature attests to the presence of powerful
and frequent magnetic flares in T Tauri stars, the geometry of the
reconnecting fields is still under debate \citep{Feigelson99}.  We
consider three alternative geometries, each with a different
implication for the efficiency of proton irradiation of disk material.

\subsubsection{Stellar surface fields \label{surface_subsec}}

Optical photometry and Doppler imaging of active regions, and Zeeman
splitting of photospheric absorption lines demonstrate that T Tauri
stars have multipolar magnetic fields like the Sun but greatly enhanced
in strength and surface coverage.  The Zeeman measurements indicate
average surface field strengths $Bf \simeq 2-4$ G, where $f$ is the
surface fraction covered with fields \citep{Valenti01}.  Study of X-ray
flare emission during eclipses of the magnetically active binary Algol
show that its emission is confined to the close vicinity of the stellar
surface \citep{Schmitt99}, but no similar direct measurement is
available for T Tauri stars.  Models of young stellar flare decays
based on radiative cooling of an instantaneously heated plasma suggest
magnetic loops much larger than the stellar radius.  But models
invoking continuous reheating are consistent with smaller loops of
order $\simeq 0.1$ R$_\star$ in extent \citep{Favata01}.

If the X-rays are produced from reconnecting multipolar fields near the
stellar surface as in the Sun then, assuming a direct ballistic
trajectory, 13\% of the emission will impact a hypothetically flat thin
disk at an oblique angle \citep[][98ff]{Hartmann98}.  Theoretical
considerations \citep{Chiang97} and direct imaging of protostellar
disks \citep{Bell97} indicate disk flaring (i.e. vertical scale height
increases with distance from the young star) is common on $>1$ A.U.
scales, so the impact fraction is likely higher.  However, energetic
protons released by reconnecting fields must follow magnetic field
lines which are likely to have a complicated geometry that may differ
substantially from the solar field.  Some protons may be released into
a closed dipole magnetosphere, while others may follow open lines that
perpendicular to the disk that confine the bipolar outflows
\citep{Shu00}.  It is unclear whether a substantial fraction of
energetic protons would be focused onto the disk.  This question may be
answered by the next generation of radio interferometers which could
image the spatial distribution of flare-generated MeV electrons (\S
\ref{geom_test_subsec}).

Two features in our $Chandra$ Orion results suggest that pre-main
sequence flares arise near the stellar surface as in main sequence
stellar flares.  First, short-duration secondary flares are seen
superposed on the long decay phase of a powerful flare in a number of
cases (see JW 470, 567 and 826 in Figure \ref{lightcurves_fig}).  This
is commonly seen in powerful solar long duration flares as different
portions of a complex active region release magnetic energy.  Second,
some of the ACIS spectra show emission lines indicative of elemental
abundance anomalies seen in stars whose emission processes are
definitely unrelated to circumstellar disks.  A considerable number of
ONC ACIS sources, including some solar analogs, show spectra suggesting
overabundance of neon and underabundance of iron.  This combination of
plasma abundance anomalies, nicknamed the `anti-FIP Effect' (First
Ionization Potential), has been seen in the X-ray spectra of flares of
magnetically active main sequence and RS CVn giant stars
\citep{Brinkman01, Gudel01}.  The reasons for these plasma abundance
anomalies are still uncertain but are thought to involve the
evaporation of chemically fractionated regions of the stellar surface
by flare energetic particles.  It is unclear whether a similar
fractionation could occur at the surface of a circumstellar disk.

\subsubsection{Star-disk fields \label{stardisk_subsec}}

Considerable evidence suggests that young stars have strong dipole
magnetic fields which extend out to the corotation radius of the
circumstellar disk \citep[see reviews by][]{Hartmann98, Feigelson99,
Shu00}.  Star-disk field lines simultaneously funnel accreting material
onto the stellar surface (this gas is responsible for T Tauri broad
optical emission lines that had been formerly attributed to a dense
stellar wind), and likely transfer stellar angular momentum from the
star to a bipolar outflow \citep[see however][]{Stassun99}.

In this case, it is reasonable that some magnetic field lines will
extend from the disk to portions of the disk that are not in corotation
with the star and will thereby suffer twisting and reconnection.  This
might occur if the disk has inhomogeneous accretion rate \citep{Shu97},
if the stellar rotation is not yet decelerated \citep{Montmerle00}, or
if field lines are coupled to the shearing layers in the Keplerian disk
\citep{Birk00}.  Numerical MHD calculations indicate that the
interaction between the stellar dipolar and disk toroidal fields is
unstable and can erupt with a violent magnetic reconnection
\citep{Hayashi96}.  In this geometry, the flare X-rays and particles
are  produced close to and above the inner disk, and spallation of
inner disk material is likely to occur in an efficient manner. This is
the geometry assumed in the meteoritic spallation calculations of
\citet{Lee98} and \citet{Gounelle01} and in the chondrule melting model
of \citet{Shu01}.

\subsubsection{Disk fields \label{disk_subsec}}

It is likely protostellar disks amplify embedded magnetic fields
through turbulence and dynamo processes, and it is possible that
violent reconnection occurs when these fields emerge at the disk
surfaces.  Ionization from cosmic rays and young stellar X-rays are
sufficient to couple the largely neutral disk gas to field lines and
induce turbulence via the magneto-rotational (Balbus-Hawley)
instability over the outer layers of the protostellar disk
\citep{Gammie96, Igea99}.  Numerical MHD models show heating and
ionization of a disk corona \citep{Miller00}, but these calculations do
not include the physics for reconnection and flares.  Detailed modeling
of flaring from magnetic fields subject to the Keplerian shear of an
accretion disk is just beginning \citep{Romanova98, Poutanen99}.

Disk magnetic fields and flares have also emerged in meteoritical
studies.  The `remanent magnetism' of chondrules in a wide range of
stony meteorites gives direct evidence for $\sim 1$ G fields in the
solar nebula \citep{Levy78}.  In addition, some models of chondrules
melting and proton spallation are based on irradiation from disk field
flares:  \citet{Levy89} consider flares in a disk corona during the
earliest evolutionary phases, while \citet{Cameron95} considers flares
at the interface between the young solar wind and a receding disk
during a later pre-main sequence phase.  In both cases, the advantage
of disk flares is to produce chondrules and radioactive isotopes {\it
in situ} at the Asteroid Belt where the parent bodies of meteorites are
now found.

\subsubsection{Observational tests of magnetic geometry
\label{geom_test_subsec}}

Two observational tests of these alternatives are feasible.  First,
some X-ray flares from young stars are sufficiently hot ($kT \geq 5$
keV) and luminous that the continuum around the K$\alpha$ transitions
of iron is seen with high signal-to-noise.  Such cases provide the
opportunity to see a fluorescent emission line reflected from neutral
iron atoms in the circumstellar disk.  This fluorescent line at 6.4 keV
can be distinguished from the $6.6-6.9$ keV thermal line complex
produced by the highly ionized iron species in the flare plasma at CCD
spectral resolution.  Double emission line complexes attributable to
hot and cold iron have been reported in two very young protostellar
systems:  infrared source R1 in the Corona Australis Coronet Cluster
using the ASCA SIS detector \citep{Koyama96} and YLW 16A in the $\rho$
Ophiuchi cloud core using the Chandra ACIS-I detector
\citep{Imanishi01}.  In both cases, the equivalent width of the two
iron lines were comparable, indicating that the disk subtends a large
solid angle when viewed by the X-ray emitting magnetic structure.

This is evidence against reconnection of magnetic fields very close to
the stellar surface (\S \ref{surface_subsec}), and favors a geometry
where the reconnection occurs above the disk in a `lamppost'
configuration (\S \ref{stardisk_subsec}-\ref{disk_subsec}).
Fluorescent iron line emission under a variety of geometries has been
investigated in the context of Seyfert galaxies and X-ray binary
systems \citep[][and references therein]{Nayakshin01}, although similar
study has not yet been conducted specifically for protostellar
systems.  As in Seyferts, the high amplitude variability of the flare
emission should be reproduced in the fluorescent line with time delays
associated with the relative geometry of the emitting and reflecting
components and the line-of-sight.   This is a potentially powerful tool
for understanding protostellar systems, but the limited signal-to-noise
around 6.4 keV provided by existing telescopes permits only limited
reverberation mapping.  \citet{Imanishi01} report that the emitting and
reflecting components in YLW 16A vary together with a delay $< 10^4$
light-seconds indicating that the flare-disk distance is  $< 15$ A.U.
Higher signal-to-noise reverberation mapping studies may be very
fruitful with planned high-throughput X-ray missions such as
Constellation-X and XEUS.

The second method for measuring magnetic geometries in young stellar
systems uses radio band Very Long Baseline Interferometry (VLBI) which
can image the distribution of transrelativistic electrons emitting
gyrosynchrotron radiation.  It is likely that the spatial distribution
of MeV protons is similar to that of MeV electrons, as both must follow
the magnetic field lines from the sites of particle acceleration. A
VLBI survey of several radio- and X-ray-bright young stellar objects in
nearby star forming regions revealed extended emission on scales of
$\sim 1$ mas or $\sim 200$ A.U. \citep{Phillips91}, but in at least one
case the emission was found to arise from both components of a close
binary \citep{Phillips96}.  There is one example of a clearly resolved
nonthermal young stellar radio source:  T Tau South, the embedded
companion to the optically bright prototype T Tau,  exhibits variable
radio emission concentrated in two jets where each jet produces
gyrosynchrotron with oppositely oriented circular polarization
\citep{Ray97}. If this geometry is ubiquitous, we might conclude that
energetic protons produced by protostellar flares are transported away
from the outer disk by the bipolar outflows.  But it is still possible
in this situation that the particles impact inner disk solids at the
base of the outflow, as in the proton irradiation model of
\citet{Shu97} and \citet{Gounelle01}.  The planned Extended Very Large
Array and Atacama Large Millimeter Array should have the sufficient
capabilities to resolve the magnetic structures containing relativistic
particles in nearby YSOs.  This requires mapping circularly polarized
sub-milliJansky continuum emission  on milliarcsecond scales.

\subsection{Can the impacting protons produce the anomalies?
\label{anom_subsec}}

A full analysis of the spallogenic effect of the energetic protons in
pre-main sequence stellar systems lies beyond the scope of this paper.
We restrict our discussion to the relationship between two recent
theoretical investigations of spallation in the solar nebula and the
proton fluences inferred to be present in the present study.

\citet{Lee98} and \citet{Gounelle01} consider the effect of a powerful
reconnection event occurring above the corotation radius of the
circumstellar disk having $\log L_x = 5 \times 10^{30}$ erg s$^{-1}$
and a luminosity in $E>10$ MeV protons of $L_p = 4.5 \times 10^{29}$
erg s$^{-1}$.  Their spallation model is applied to proto-CAI material
near the disk corotation radius and produces observed abundances of
several short-lived nuclides: $^{10}$Be, $^{26}$Al, $^{41}$Ca,
$^{53}$Mn and $^{138}$La but not $^{60}$Fe.  They treat self-shielding
in CAIs $\geq 1$ mm in size, which is important to reduce
overproduction of $^{41}$Ca.  Their model requires specification of
flaring magnetic field geometry (a `lamppost' loop over the inner
disk),  $^3$He enrichment of flare particles\footnote{The overabundance
of neon seen in the X-ray spectra of powerful stellar flares
\citep[][, \S \ref{surface_subsec}]{Brinkman01, Gudel01} and the
association of neon with helium abundances in solar energetic particles
\citep{Dwyer01} support this idea.  If it is due to a chemical
fractionization by first ionization potential, then helium and neon
abundances should be correlated in flare plasmas which are the likely
source of accelerated particles as helium and neon have the highest
first ionization potentials of all elements.  Recall that helium lines
cannot be detected in the X-ray plasmas because it is fully ionized.},
and a mechanism for dispersing the CAIs from the inner to the mid-disk
region by temporary entrainment into the bipolar outflow.

\citet{Goswami01} study the spallogenic production of $^{26}$Al,
$^{36}$Cl, $^{41}$Ca and $^{53}$Mn by solar enegetic particles in the
mid-disk region.  They consider reactions on both gas- and solid-phase
atoms using energetic protons and $^4$He (but not $^3$He) with
self-shielding in the larger particles.  The cause and geometry of the
flare event is not specified. Proton fluxes are normalized to a
time-averaged level of  $f_p (E>10~{\rm MeV}) \simeq 100$ protons
cm$^{-2}$ s$^{-1}$ observed at 1 A.U. from the contemporary Sun.  They
find that meteoritic abundances of $^{41}$Ca, $^{36}$Cl and $^{53}$Mn
require $5 \times 10^3$ to $10^4-$fold enhancement of the proton flux
for $5 \times 10^5$ to several million years, while production of
$^{26}$Al  requires enhancement factor of almost $10^5$.

We find that both of these calculations of a solar nebula origin of
meteoritic short-lived nuclei are fully consistent with the flare
proton levels we infer are present during the first million years of
the solar system. The X-ray and proton luminosities assumed by
\citet{Gounelle01} to explain the observed abundances of several
nuclides are nearly identical to the values we find from ONC solar
analogs (\S \ref{inferred_sec}).  The proton enhancement factors and
irradiation durations required by \citet{Goswami01} similarly
consistent with the $10^5$ enhancement over $\sim 1$ My that we infer
from the Chandra ONC results (\S \ref{inferred_sec}).  Recall that
their enhancement factors apply to particle fluences observed at a
distance of 1 A.U. from the proton acceleration site, so that lower
enhancement factors would be needed if the flaring magnetic structures
were closer to the irradiated disk material, as in \S
\ref{stardisk_subsec}-\ref{disk_subsec}.  Our observational findings on
ONC solar analogs and these calculations of meteoritic isotopic
anomalies from spallation by locally produced particles are thus
completely compatible completely compatible with each other, except
for unfavorable magnetic field geometries.

\section{Conclusions}

This study addresses long-standing puzzles regarding the origin of high
energy processes in the solar nebula, as evidenced by petrologic,
isotopic and cosmic ray tracers in ancient pristine meteorites. Based
on the general knowledge of enhanced X-ray flaring in pre-main sequence
stars available a decade ago, \citet{Feigelson91} concluded that X-ray
and radio flare studies of T Tauri stars "have a clear contribution to
these debates concerning meteoritic properties: {\it invoking high
levels of energetic particle or radiation fluxes associated with
magnetic activity in the early Sun is no longer ad hoc}".  The
$Chandra$ measurements of flaring in Orion pre-main sequence solar
analogs presented here confirm this statement and provide details that
were not available earlier.

We have established that virtually all analogs of the $\leq 1$ My old
pre-main sequence Sun exhibited X-ray flares that are $10^{1.5}$ times
more powerful and $10^{2.5}$ times more frequent than the most powerful
flares seen on the contemporary Sun (\S \ref{inferred_sec}).  The
inferred energetic proton flux is increased by an additional factor of
$10^1$ above the X-ray flux, although the physical mechanism underlying
this enhancement factor is uncertain. We infer that the proton fluence
in the early Sun is about $10^5$ times the time-averaged levels
produced by the contemporary Sun.

The line of reasoning linking the ONC X-ray measurements to solar
nebular isotopic anomalies has several steps.  The completeness of our
sample and ubiquity of strong X-ray flaring during the first My (though
a decline in activity appears in some stars during the $1-10$ My
period, \S \ref{evol_subsec}) argues strongly that the early Sun
produced such flares (\S \ref{Ori_Sun_sec}).  There are also strong
reasons, based on YSO radio gyrosynchrotron measurements and analogy
with solar flare physics, to believe that the ONC X-ray flares are
accompanied by high proton fluences (\S \ref{protons}).  The geometry
of the reconnecting magnetic fields and consequence fraction of flare
protons that impact the disk is uncertain, particularly at distant
locations in the Asteroid Belt where meteorites parent bodies are now
found (\S \ref{disk_sec}).  However, the report of the 6.4 keV iron
fluorescent line in protostellar spectra strongly suggests that the
X-rays, and by reasonable inference the protons, efficiently impact the
disk.  Unless the flare geometries are unfavorable, we find the
inferred proton flux on the disk is sufficient to produce the observed
CAI abundances of several important short-lived radioactive isotopes
(\S \ref{anom_subsec}).

Our study thus strengthens the astronomical foundation for local proton
spallation models of meteoritic isotopic anomalies.  However, it
cannot not establish the entire theory.  For example, it is not clear
that a local spallation model can account for meteoritic $^{60}$Fe
\citep{Lee98} or for the evidence that the abundance of $^{53}$Mn
was greater in the outer solar nebula than the inner nebula 
\citep{Shukolyukov00}.  It seems plausible that {\it both} external
seeding and internal spallation may have been sources of short-lived 
nuclides in the solar nebula.  \citet{Meyer00} also emphasize that
several extinct radioactive isotopes in meteorites can be attributed to 
ongoing nucleosynthesis in the Galaxy.  

Although our attention here has been principally directed towards the
isotopic anomalies, the findings here may similarly important
implications for the other meteoritic characteristics demanding high
energy processes. Our results clearly support the extensive evidence
for a flare spallogenic origin of excess $^{21}$Ne and particle tracks
in gas-rich meteoritic silicate grains \citep[e.g.][]{Caffee87,
Woolum93, Caffee01}.  The contribution of flare energetic particles to
disk and outflow ionization might be added to existing calculations
treating Galactic cosmic ray and stellar X-ray ionization
\citep[e.g.][]{Gammie96, Igea99}. The effects of flare radiation and
shocks on disk solids, including the possibility of flash melting the
chondrules, have been considered in detail in only limited contexts
\citep{Levy89, Shu01}.  More extensive calculations of such effects
based on various possible magnetic geometries may now be warranted.

\acknowledgments We thank members of the ACIS Orion research team $-$
Pat Broos, Jim Gaffney, Leisa Townsley, Yohko Tsuboi (PSU) and Lynne
Hillenbrand (Caltech) $-$ for their many efforts. The paper greatly
benefited from the thoughtful commentary of the principal referee, A.
G. W.  Cameron (Harvard/Arizona-LPL).  Valuable comments on the
manuscript were also provided by Donald Clayton (Clemson), Terry Forbes
(UNH), Alfred Glassgold (NYU/UCB), Martin Gounelle (Natural History
Museum, London), Thierry Montmerle (Saclay) and Sienny Shang (UCB).
This work was principally funded by NASA contract NAS8-38252 supporting
the Chandra ACIS team (G.\ Garmire PI).

\newpage

\appendix \section{Comparison with contemporary solar flares}

The powerful X-ray flares we detect in ONC solar analogs closely
resemble the X-ray emission from relatively rare powerful solar
flares.  Such flares constitute long duration events (10-20 hours) in
the soft X-ray band and are associated with gradual hard burst events
(GHXs) with hardening in the X-ray band from roughly 40 to 200 keV.
These events are also associated with type II and IV bursts in the
metric radio band, coronal mass ejections (CMEs), and prompt solar
energetic particle events (SEPs) in measurements of the interplanetary
medium. These properties have been nicknamed the Big Flare Syndrome
\citep{Kahler82} because small to moderate events often lack one or
more of these features.

While their physics is very complex and still debated, successful
models for many properties of solar eruptive events have been
developed.  Their power and duration require continuing release of
energy previously stored in magnetic fields at the solar surface.  In
many models, a flux rope rises in an arcade of field loops, resulting
in field reconnection to a simpler geometry and the ejection of
magnetic flux at velocities of 1000-2000 km s$^{-1}$ \cite[][ \S
11.1]{Priest00}.  The outer loops of the arcade are filled with $10^7$
K plasma, while the inner loops have $10^4$ plasma appearing as two
ribbons of H$\alpha$ emission (hence the nomenclature `two-ribbon
flare'). Both radiative and conductive cooling may be important in the
heated loops.  Some gas flows downward onto the surface, evaporating
chromospheric material which then flows upwards along other field
lines; this may be responsible for unusual elemental abundances seen in
some solar and stellar X-ray flare plasmas.

The production of energetic (MeV) particles is less well understood but
may involve several processes acting alone or in consort: direct
acceleration by strong electric fields between different magnetic
regions of the flare, or acceleration in MHD waves, shocks,
instabilities or turbulence \citep[][, \S 13]{Miller97, Priest00}.
Particles can be accelerated both on small scales during the impulsive
phase of a flare, on larger scales within the coronal mass ejection, or
on very large scales in corotating interaction regions of the solar
wind. During the brief impulsive flare, gamma-ray line and continuum
emission, neutrons from nuclear collisions, microwave gyrosynchrotron,
and hard X-ray bremsstrahlung can be present.  Impulsive particle
acceleration appears to be very efficient with $\sim 10^{31}$ erg
produced in each of energetic electrons, protons and ions in an X-class
flare.  Protons can acquire up to GeV energies.  However, many of these
particles remain trapped in solar fields and are not ejected into the
interplanetary medium.

During $1989-1997$ (solar cycle 22), a period of about 3000 days, the
{\it Geosynchronous Operational Environmental Satellite (GOES) 7}
satellite detected about 20 solar flares with $28.0 \leq \log
L_x(peak)  \leq 29.0$ erg s$^{-1}$ in the $1.6-12$ keV ($1-8$ \AA) band
\citep{Sammis00}.  The recurrence rate of these maximally X-ray
luminous solar flares, which are $\simeq 100$ times weaker than the
typical ONC flare is about one flare per year ($3 \times 10^7$ s).

\citet[][updated in Reedy 2001]{Reedy96} has obtained an understanding
of the fluence of solar enegetic particles at 1 A.U.  He combines
direct spacecraft measurements of solar energetic particles from
1965-91 (solar cycles $20-22$) with indirect measurements from $^{14}$C
in tree ring data and abundances of spallogenic nuclides in lunar
rocks.  The results are all consistent with a constant time-averaged
omnidirectional proton flux over the past $10^1-10^6$ yr of $f_p \simeq
200$ protons cm$^{-2}$ s$^{-1}$ at 1 A.U. for energies $E > 10$ MeV.
Most of these protons are produced in the most powerful flares as the
integrated frequency of events with fluence  $F$ is given
by\footnote{This relation differs somewhat from the differential
distribution $dN/dE \propto E^{-1.15}$ reported by
\citet{vanHollebeke75}.  While indices ranging from -1.15 to -1.4 have
been reported for the distribution of solar proton events, they are
always substantially flatter than the -1.8 index found for the
distribution of X-ray peak fluxes \citep{Hudson91}.} $N(F>F_o) \simeq 1
(F_o/10^9 {\rm protons~cm}^{-2})^{-0.4}$ yr$^{-1}$ over the range $10^7
< F_o < 10^{10}$ protons cm$^{-2}$.

Reedy finds a steep cutoff in solar particle events with fluences above
$10^{10}$ protons cm$^{-2}$.  The reasons for this are unknown.  We
assume here that this limit applies only to contemporary solar flares,
and that the proton fluences associated with young stellar flares
exhibiting X-ray luminosities far above those seen in the Sun today can
considerably exceed this solar limit.

It can be useful to examine the behavior of individual powerful solar
flares as analogs for ONC flares.  Consider, for example, a typical
strong solar flare and proton event like the 23-24 March 1991 flare
\citep[SOLTIP Interval 1; ][see details in Watanabe 1995]{Shea96}.  The
$GOES-7$ lightcurve shows the Sun's X-ray luminosity rise over $\simeq
4$ hours from $8 \times 10^{24}$ to a peak of $1 \times 10^{26}$ erg
s$^{-1}$ in the $1.6-12$ keV band.  The X-ray decay over $\simeq 20$
hours is accompanied by several $\simeq 1$ hr secondary flares with
peak luminosities comparable to the principal flare.  The flare was
followed by an increase in $E>1$ MeV proton fluence from $10^1$ to
$10^{5.5}$ protons cm$^{-2}$ s$^{-1}$ ster$^{-1}$ at 1 A.U.  over the
next day, which produced a major geomagnetic storm, terrestrial
electromagnetic anomalies, and a prominent decrease in the Galactic
cosmic ray intensity.  Note, however, that the flare on the previous
day, 22 March, which was an intense gamma-ray and relativistic neutron
emitter, produced a brief impulsive X-ray spike but not a long duration
soft X-ray event.

An exceptionally powerful group of solar flares occurred over two weeks
in June 1991 (SOLTIP Interval 2).  Several received the highest
possible X-ray classification $>$X12.  The 1 June 1991 flare, for
example, produced:  (a) $10^{32}-2 \times 10^{33}$ erg in X-rays (the
exact value is unknown due to saturation of the {\it GOES} detectors);
(b) more than $10^{32}-10^{34}$ erg injected (but not necessarily
released into the interplanetary medium) in $>20$ keV electrons; (c)
$\sim 10^{33}$ erg in MeV nucleons; (d) a powerful flux of neutrons;
and (e) strong shocks in the interplanetary medium driven by
flare-induced $^3$He-enriched coronal mass ejection \citep{Kane95,
Watanabe95, Ramaty97, Murphy99, Clayton00}.  The peak X-ray luminosity
in the {\it GOES} $1.6-12$ keV band is estimated to be $\log L_x \simeq
10^{28.5}$ erg s$^{-1}$.   No major proton flare was seen at the Earth
from this event as the interplanetary magnetic field lines carrying the
particles fortuitously missed the Earth.

\newpage

\begin{deluxetable}{rccrrrrccrrrcrccrr}
\rotate
\tabletypesize{\scriptsize}
\tablewidth{0pt}
\tablecolumns{18}
\tablecaption{X-ray properties of pre-main sequence solar analogs in the ONC 
\label{analogs_tab}}

\tablehead{
\multicolumn{2}{c}{ACIS source} &&
\multicolumn{5}{c}{Optical properties} &&
\multicolumn{9}{c}{X-ray properties} \\ 
\cline{1-2} \cline{4-8} \cline{10-18}

\colhead{Src \#} &
\colhead{CXOONC J} &&
\colhead{JW} &
\colhead{$\log L_*$} &
\colhead{$A_V$} &
\colhead{$\log t$} &
\colhead{Disk} &&
\colhead{$C_{xtr}$} &
\colhead{$CR_1$} &
\colhead{$CR_2$} &
\colhead{Var cl} &
\colhead{$\log N_H$} &
\colhead{$kT$} &
\colhead{S.F.} &
\colhead{$\log L_x$} &
\colhead{$\log L_x/L_*$} \\

&&&&
\colhead{($L_\odot$)} &
\colhead{(mag)} &
\colhead{(yr)} &&&&
\multicolumn{2}{c}{(cts ks$^{-1}$)} &&
\colhead{(cm$^{-2}$)} &
\colhead{(keV)} &&
\colhead{(erg s$^{-1}$)} & 
}
\startdata
\nodata&   \nodata       &&  62~ &  0.4~ & 5.3 &    6.2 &  $-$  && $<$30 
&\nodata&\nodata&\nodata&\nodata &\nodata&\nodata &$<$29.5~ & $<$-4.5~~ \\
    43 & 053451.5-052512 && 123~ &  0.3~ & 3.1 &    6.4 &   +   &&   436 &  4.8  &  5.9 & Const  &   
21.4~ & 1.4  & $\star$ & 29.8~ & -4.1~~ \\
   109 & 053500.4-052513 && 198~ & -0.4~ & 0.6 &    6.6 &   +   &&   586 &  8.5  &  5.6 & Pos fl &   
21.3~ & 3.2  & \nodata & 30.1~ & -3.1~~ \\
   179 & 053505.7-052418 && 278~ & -0.1~ & 1.7 &    6.2 &   +   &&    63 &  0.4  &  1.2 & LT Var &   
21.5~ & 4.2  & \nodata & 29.2~ & -4.3~~ \\
   185 & 053506.2-052202 && 286~ &  0.0~ & 3.9 &    6.1 &   +   &&  1182 & 15.8  & 13.7 & Const  &   
22.1~ & 2.0  & $\star$ & 30.5~ & -3.1~~ \\
   218 & 053508.3-052829 && 320~ & -0.1~ & 3.6 &    6.9 &   +   &&    40 &  0.6  &  0.2 & Pos fl &   
22.4~ & 2.3  & $\star$ & 29.1~ & -4.4~~ \\
   229 & 053508.8-053149 && 328~ &  0.1~ & 1.2 &    6.1 &   +   &&   277 &\nodata&  7.6 & Flare  &   
20.9~ & 4.5  & \nodata & 30.4~ & -3.3~~ \\
   230 & 053508.9-052959 && 330~ &  0.5~ & 0.9 &    5.9 &   $-$ &&  1724 & 27.9  & 10.7 & Flare  
&\nodata & 0.8/2& $\star$ & 30.5~ & -3.6 ~~\\
   253 & 053510.2-052321 && 341~ & -0.8~ & 0.0 &    7.5 &\nodata&&   754 &  9.1  &  9.9 & Pos fl &   
22.1~ & 2.2  & $\star$ & 30.3~ & -2.5~~ \\
   284 & 053511.4-052602 && 365~ &  0.9~ & 2.3 &    5.5 &   +   &&  2873 & 51.4  & 13.8 & Pos fl &   
21.4~ & 1.9  & \nodata & 30.8~ & -3.7~~ \\
   302 & 053511.9-051926 && 370~ &  0.7~ & 5.5 &    5.6 &   +   &&   445 &  5.0  &  6.7 & Const  &   
21.8~ & 2.6  & \nodata & 30.0~ & -4.3~~ \\
   307 & 053511.9-052033 && 373~ &  1.0~ & 5.9 & $<$5.5 &   +   &&  1378 & 15.1  & 19.8 & Pos fl &   
21.9~ & 2.0  & \nodata & 30.5~ & -4.1~~ \\
   443 & 053514.9-052239 && 454~ &  0.8~ & 2.4 & $<$5.5 &   +   &&  2095 & 13.3  & 41.2 & Flare  &   
21.5~ & 2.3  & $\star$ & 30.7~ & -3.7~~ \\
   448 & 053514.9-052159 && 457~ &  0.5~ & 2.1 &    5.6 &   +   &&  1735 & 17.2  & 26.6 & Flare  &   
21.5~ & 2.0  & \nodata & 30.5~ & -3.6~~ \\
   457 & 053515.1-052254 && 463~ &  0.8~ & 2.7 & $<$5.5 &\nodata&&   193 &  1.8  &  3.6 & Pos fl &   
21.6~ & 5.5  & \nodata & 29.7~ & -4.7~~ \\
   468 & 053515.3-052215 && 470~ &  0.9~ & 3.8 & $<$5.5 &   +   &&  1230 & 21.8  &  8.6 & Flare  &   
22.0~ & 3.6  & \nodata & 30.6~ & -3.9~~ \\
   501 & 053515.8-052322 && 488a &  0.1~ & 0.0 &    6.7 &   +   &&   184 &  1.8  &  2.9 & Const  
&$<$20.0~ &$>$10 & $\star$ & 29.5~ & -4.2~~ \\
   520 & 053516.0-052036 && 504b &  0.2~ & 2.7 &    6.0 &\nodata&&  1020 & 16.9  &  7.2 & Pos fl &   
21.4~ & 2.0  & \nodata & 30.2~ & -3.6~~ \\
   582 & 053517.2-052131 && 544~ &  1.3~ & 2.9 & $<$5.5 &  $-$  &&  1909 & 15.2  & 34.6 & Flare  &   
21.5~ & 2.4  & \nodata & 30.6~ & -4.3~~ \\
   590 & 053517.3-052400 && 549~ &  0.4~ & 6.2 &    5.7 &   +   &&    56 &  0.6  &  0.8 & Const  &   
22.1~ & 2.4  & $\star$ & 29.4~ & -4.6~~ \\
   602 & 053517.5-052256 && 553a &  0.9~ & 3.6 & $<$5.5 &\nodata&&   456 &  3.4  &  8.1 & LT Var &   
21.6~ & 3.3  & \nodata & 29.4~ & -5.1~~ \\
   604 & 053517.5-051740 && 550~ &  1.1~ & 8.1 & $<$5.5 &   +   &&  3636 & 50.4  & 37.0 & Posfl 
&\nodata &1/2.4 & \nodata & 31.1~ & -3.6~~ \\
   626 & 053517.9-052245 && 567~ &  1.5~ & 1.5 & $<$5.5 &  $-$  &&  9016 & 24.0  &212.5 & Flare  &   
21.5~ & 5.8  & \nodata & 31.7~ & -3.4~~ \\
   662 & 053518.6-052313 && 596~ &  0.7~ & 3.8 &    5.7 &   +   &&  2274 & 45.6  &  3.9 & Flare  
&\nodata & 0.1/4& \nodata & 30.9~ & -3.4~~ \\
   664 & 053518.7-052256 && 598a &  0.8~ & 4.2 &    5.6 &\nodata&&   456 &  3.5  &  8.4 & Pos fl &   
22.0~ & 2.1  & \nodata & 30.1~ & -4.3~~ \\
   675 & 053518.9-052052 && 605~ & -0.8~ & 1.3 &    7.2 &   +   &&   242 &  2.2  &  4.0 & Pos fl &   
21.9~ & 2.6  & \nodata & 29.9~ & -2.9~~ \\
   676 & 053518.9-051613 && 601~ &  0.1~ & 0.5 &    6.4 &  $-$  &&   993 & 13.3  & 10.7 & Flare  
&$<$20.0~ & 1.1  & $\star$ & 30.1~ & -3.6~~ \\
   682 & 053519.0-052349 && 607~ &  0.1~ & 2.5 &    6.8 &   +   &&    28 &  0.4  &  0.2 & Const  
&$<$20.0~ &$>$10 & \nodata & 28.8~ & -4.9~~ \\
   751 & 053520.8-052121 && 662~ & -0.5~ & 1.0 &    6.9 &   +   &&    17 &  0.2  &  0.2 & Const  &   
22.2~ & 1.0  & \nodata & 28.5~ & -4.6~~ \\
   790 & 053521.8-052354 && 698~ &  0.5~ & 1.0 &    5.9 &   +   &&   202 &  4.6  &  2.4 & Flare  &   
21.6~ & 2.2  & \nodata & 29.6~ & -4.5~~ \\
   805 & 053522.1-052424 && 707~ &  0.7~ & 2.3 &    5.8 &\nodata&&   609 &  7.4  &  7.8 & Pos fl &   
21.8~ & 2.1  & \nodata & 30.1~ & -4.2~~ \\
   808 & 053522.3-052029 && 706~ &  0.4~ & 0.7 &    6.0 &  $-$  &&   652 &  9.6  &  6.0 & LT Var 
&$<$20.0~ & 1.0  & $\star$ & 29.9~ & -4.1~~ \\
   814 & 053522.4-052201 && 712~ &  0.3~ & 4.0 &    6.1 &   +   &&   531 &  7.6  &  5.5 & Pos fl &   
21.8~ & 2.5  & \nodata & 30.1~ & -3.8~~ \\
   848 & 053523.6-052332 && 738~ & -0.2~ & 1.5 &    6.4 &\nodata&&  2906 & 56.0  & 14.1 & Flare  &   
21.9~ & 2.8  & \nodata & 30.9~ & -2.5~~ \\
   889 & 053525.0-052258 && 766a &  0.6~ & 4.4 &    5.9 &\nodata&&   524 &  5.1  &  8.3 & Pos fl &   
21.8~ & 2.2  & $\star$ & 30.1~ & -4.1~~ \\
   890 & 053525.0-052346 && 769~ &  0.5~ & 1.9 &    5.5 &  $-$  &&   719 & 10.7  &  6.7 & Flare  &   
21.1~ & 1.5  & $\star$ & 30.0~ & -4.1~~ \\
   929 & 053527.3-052336 && 810~ &  0.5~ & 3.9 &    5.6 &   +   &&  1127 &  7.0  & 22.7 & Flare  &   
21.6~ & 2.6  & \nodata & 30.4~ & -3.7~~ \\
   949 & 053528.2-052458 && 826~ & -0.6~ & 0.0 &    6.8 &   +   &&  2586 &  8.3  & 61.3 & Flare  &   
21.9~ & 2.6  & $\star$ & 30.8~ & -2.2~~ \\
   969 & 053529.9-053252 && 847~ &  0.8~ & 0.5 &    5.6 &  $-$  &&   608 &\nodata& 17.0 & Pos fl &   
21.0~ & 1.6  & \nodata & 30.6~ & -3.8~~ \\
   990 & 053531.2-052340 && 868~ &  0.2~ & 2.1 &    6.0 &   +   &&   159 &  1.1  &  1.0 & Const  &   
21.4~ & 2.7  & \nodata & 29.5~ & -4.3~~ \\
  1022 & 053534.9-052914 && 907~ &  0.2~ & 1.4 &    6.3 &  $-$  &&  1138 &  9.8  & 17.4 & LT Var 
&$<$20.0~ & 1.1  & $\star$ & 30.2~ & -3.6~~ \\
  1025 & 053535.3-052127 && 911~ &  0.2~ & 1.2 &    6.0 &   +   &&   940 & 11.7  & 10.1 & LT Var &   
21.0~ & 1.3  & $\star$ & 30.1~ & -3.7~~ \\
\nodata&    \nodata      && 991~ &  0.0~ &     &    6.8 &  $-$  && $<$50 
&\nodata&\nodata&\nodata&\nodata &\nodata&\nodata&$<$29.6~ & $<$-4.0~~ \\  
\enddata
\end{deluxetable}

\clearpage
\newpage

\clearpage
\newpage

\clearpage
\newpage

\clearpage
\newpage

\begin{figure}
\centering
  \begin{minipage}[t]{1.0\textwidth}
  \centering
\includegraphics[scale=0.30]{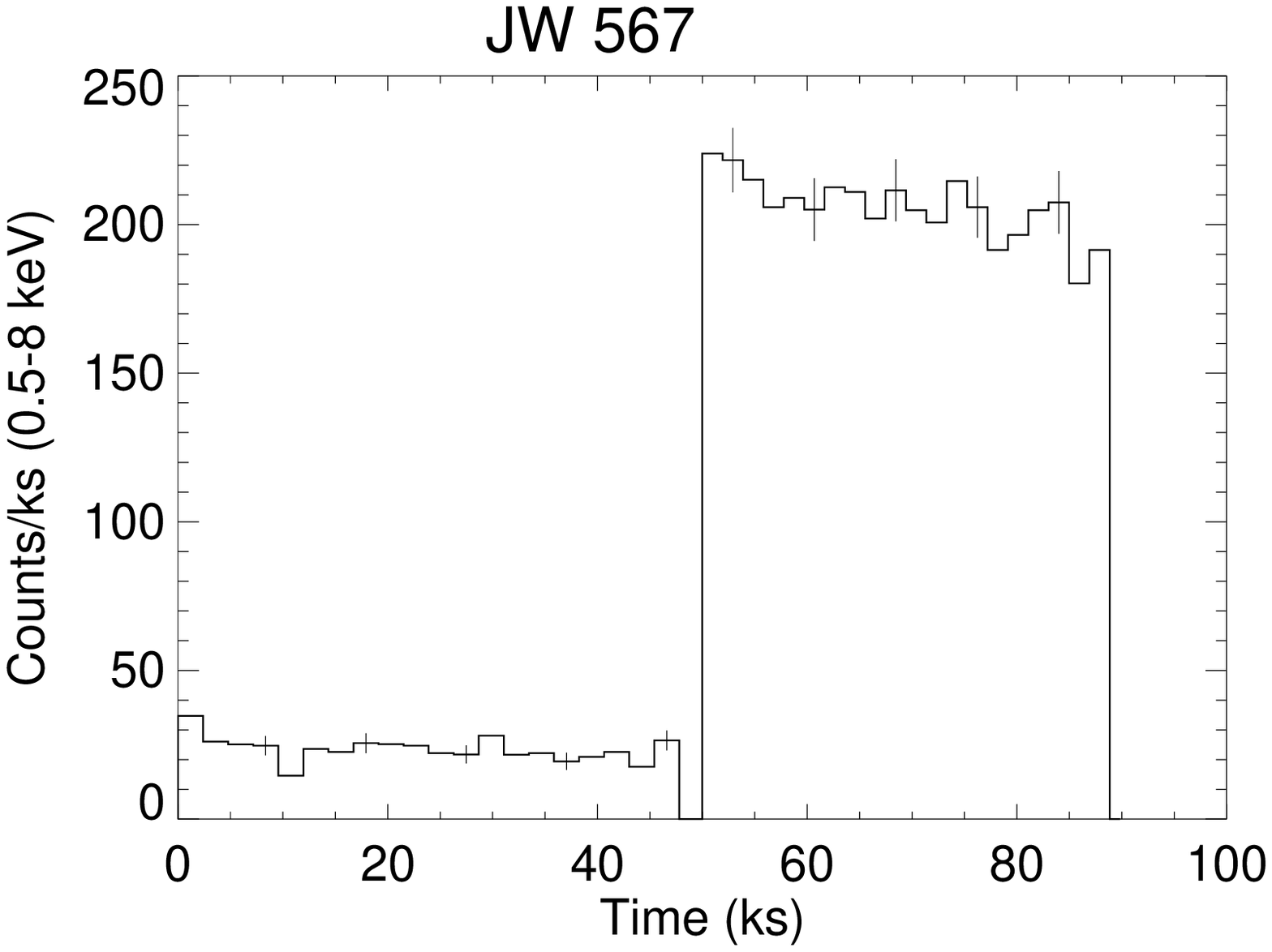}
\includegraphics[scale=0.30]{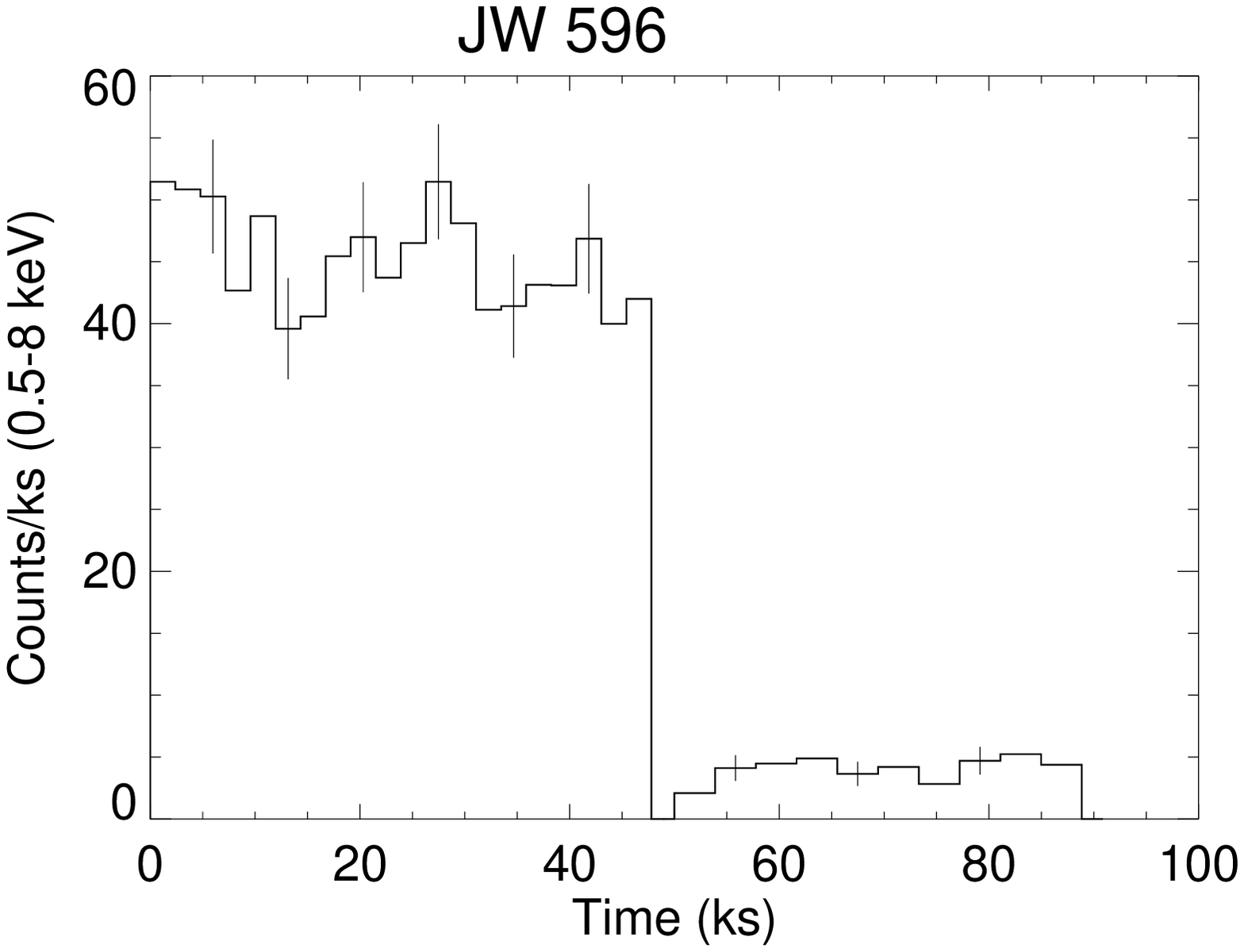}
\includegraphics[scale=0.30]{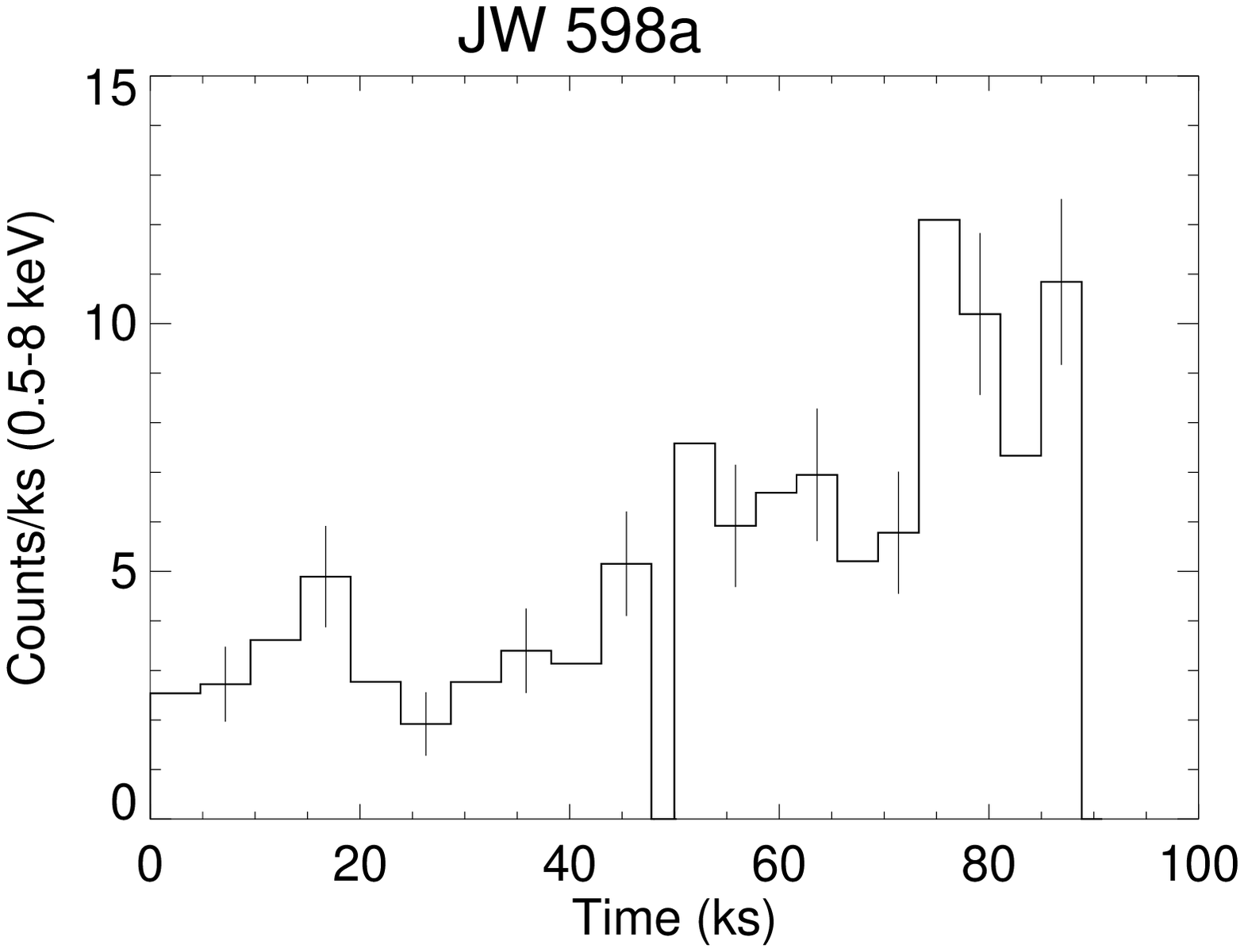}
  \end{minipage} \\ [0.3in]
  \begin{minipage}[t]{1.0\textwidth}
  \centering
\includegraphics[scale=0.30]{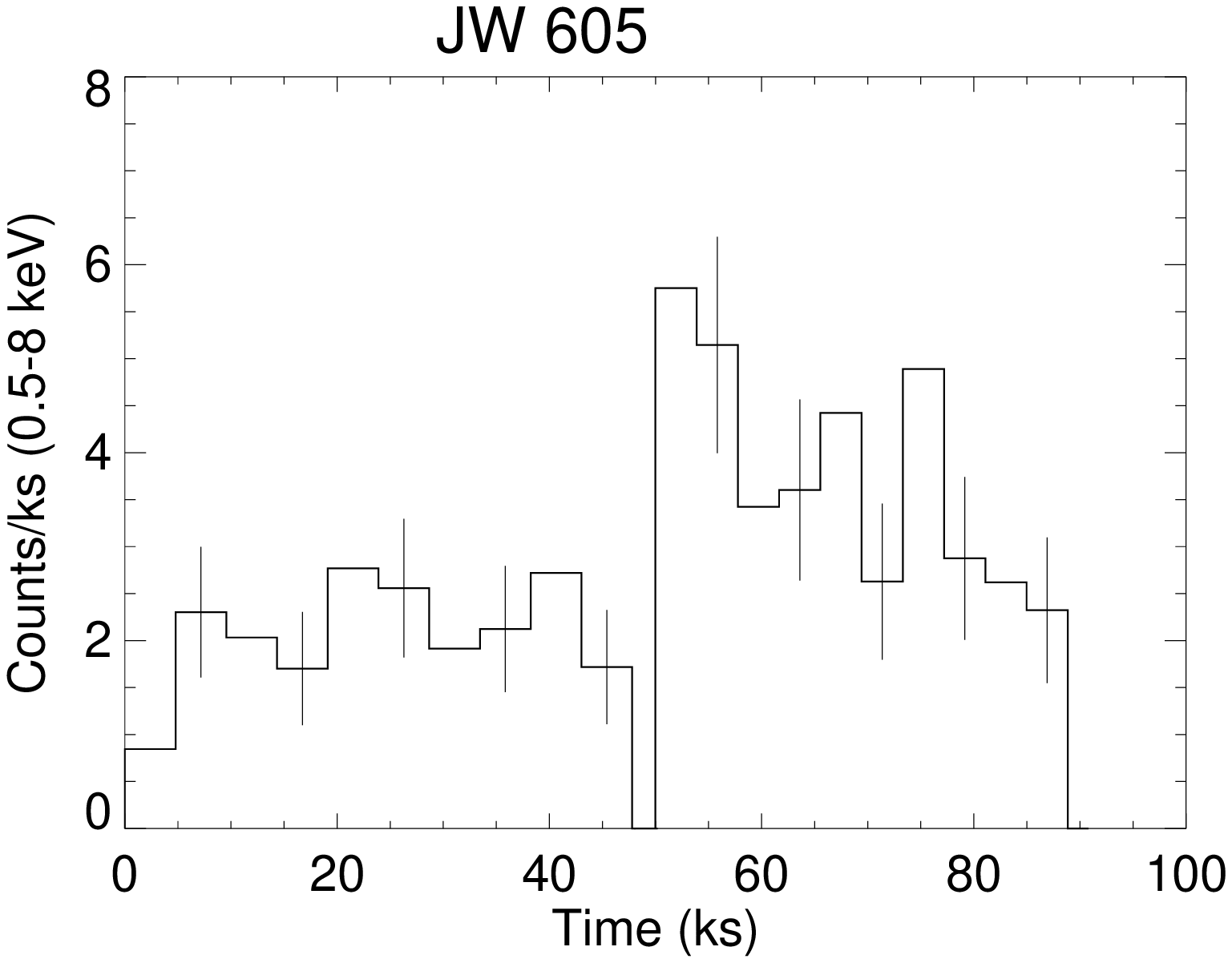}
\includegraphics[scale=0.30]{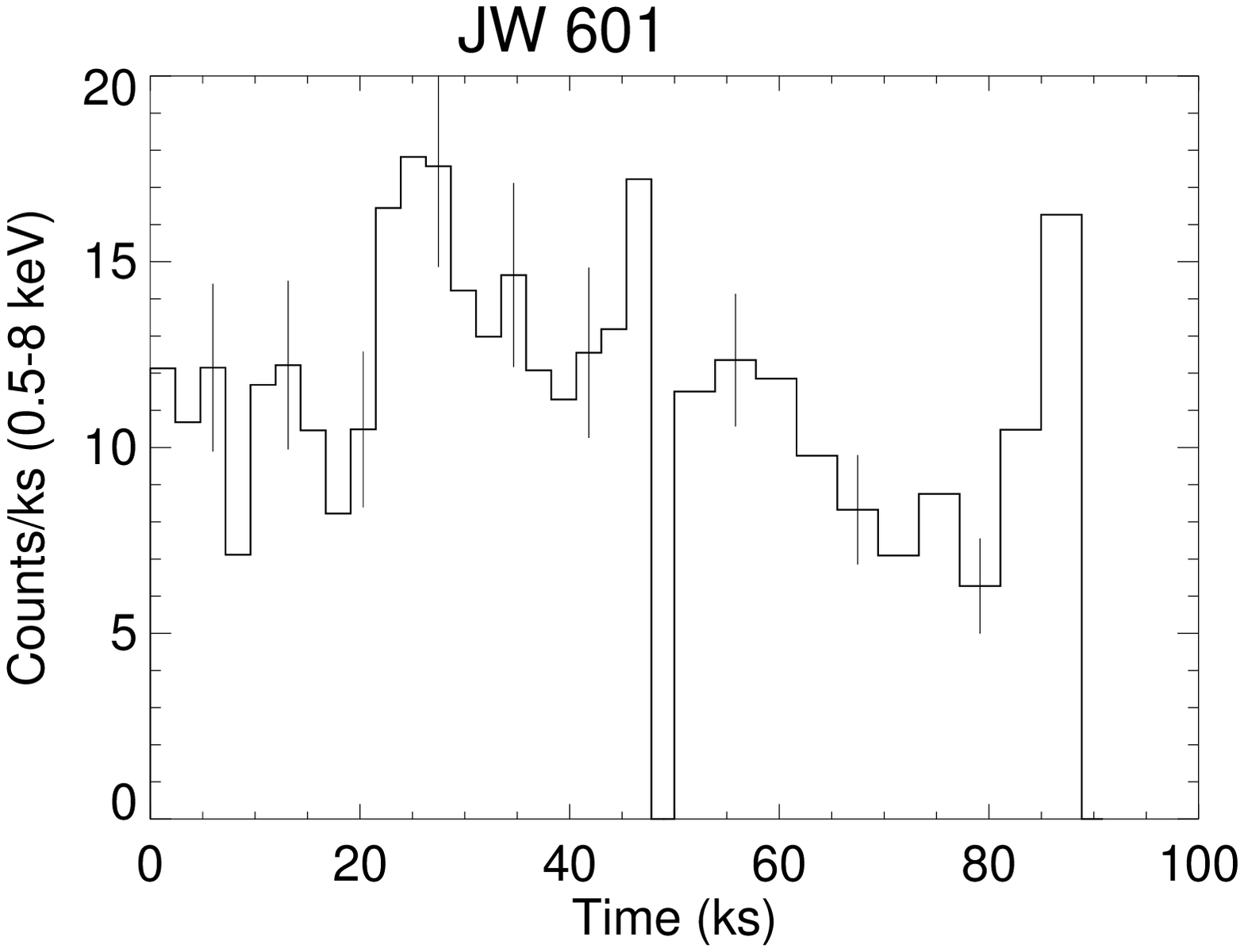}
\includegraphics[scale=0.30]{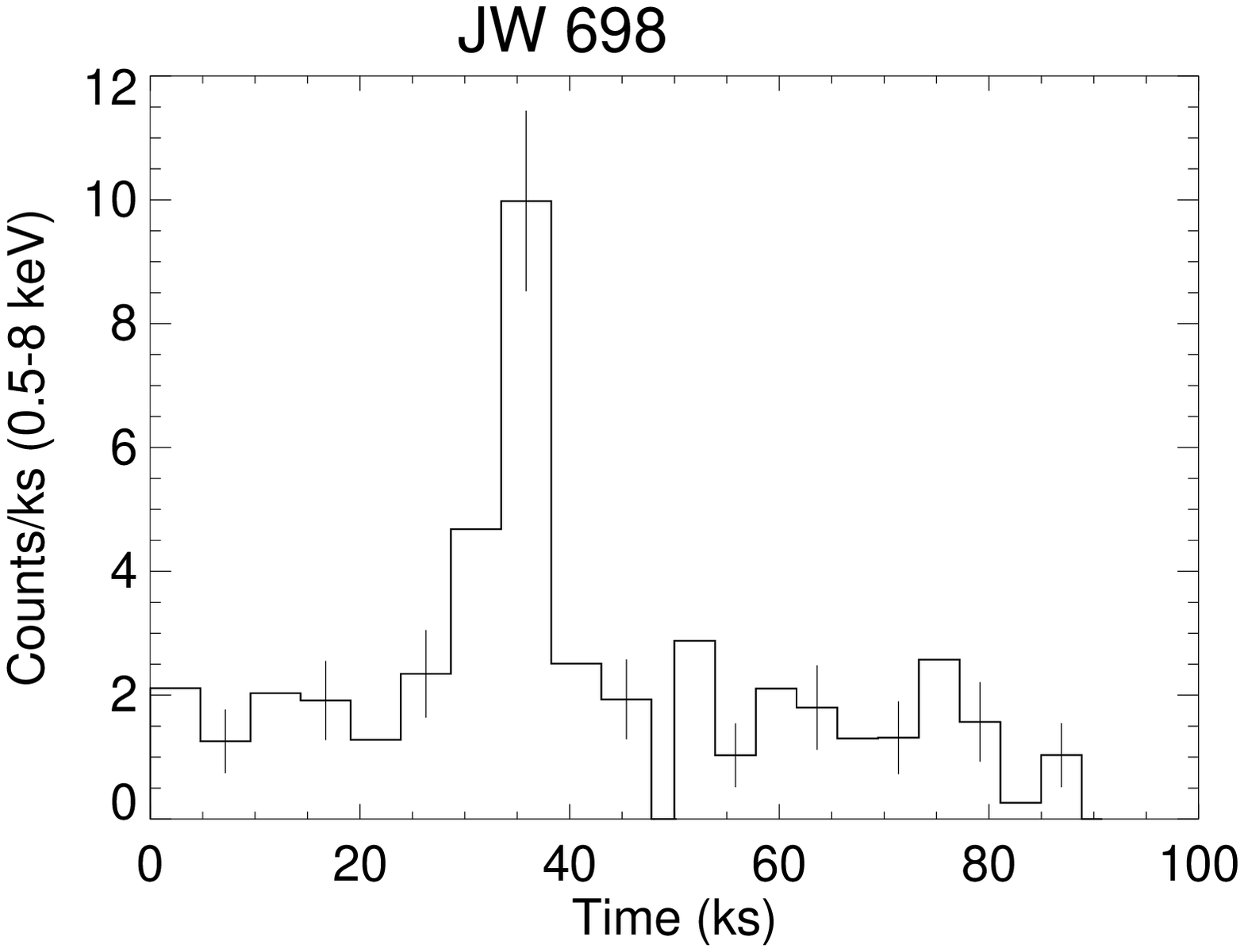}
  \end{minipage} \\ [0.3in]
  \begin{minipage}[t]{1.0\textwidth}
  \centering
\includegraphics[scale=0.30]{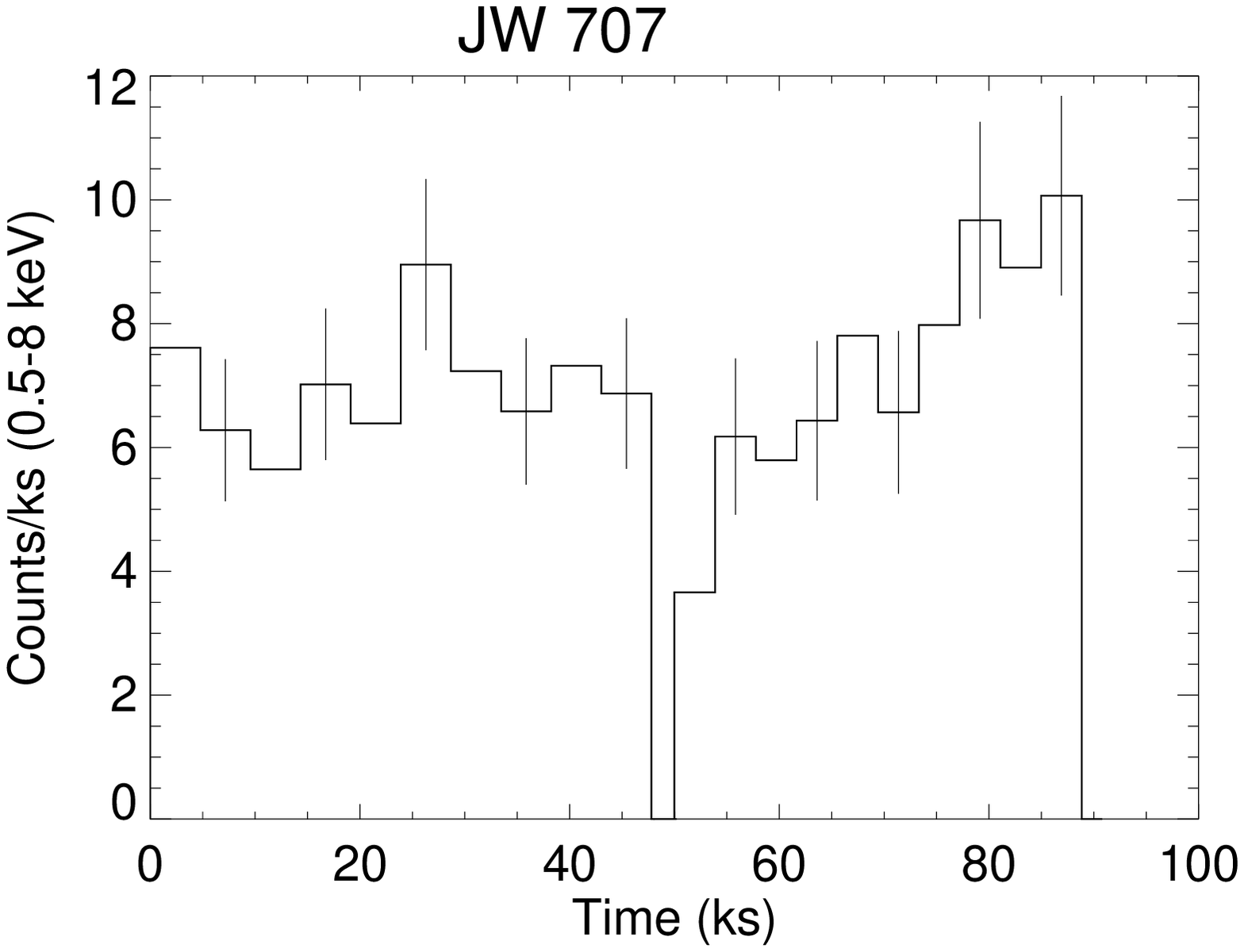}
\includegraphics[scale=0.30]{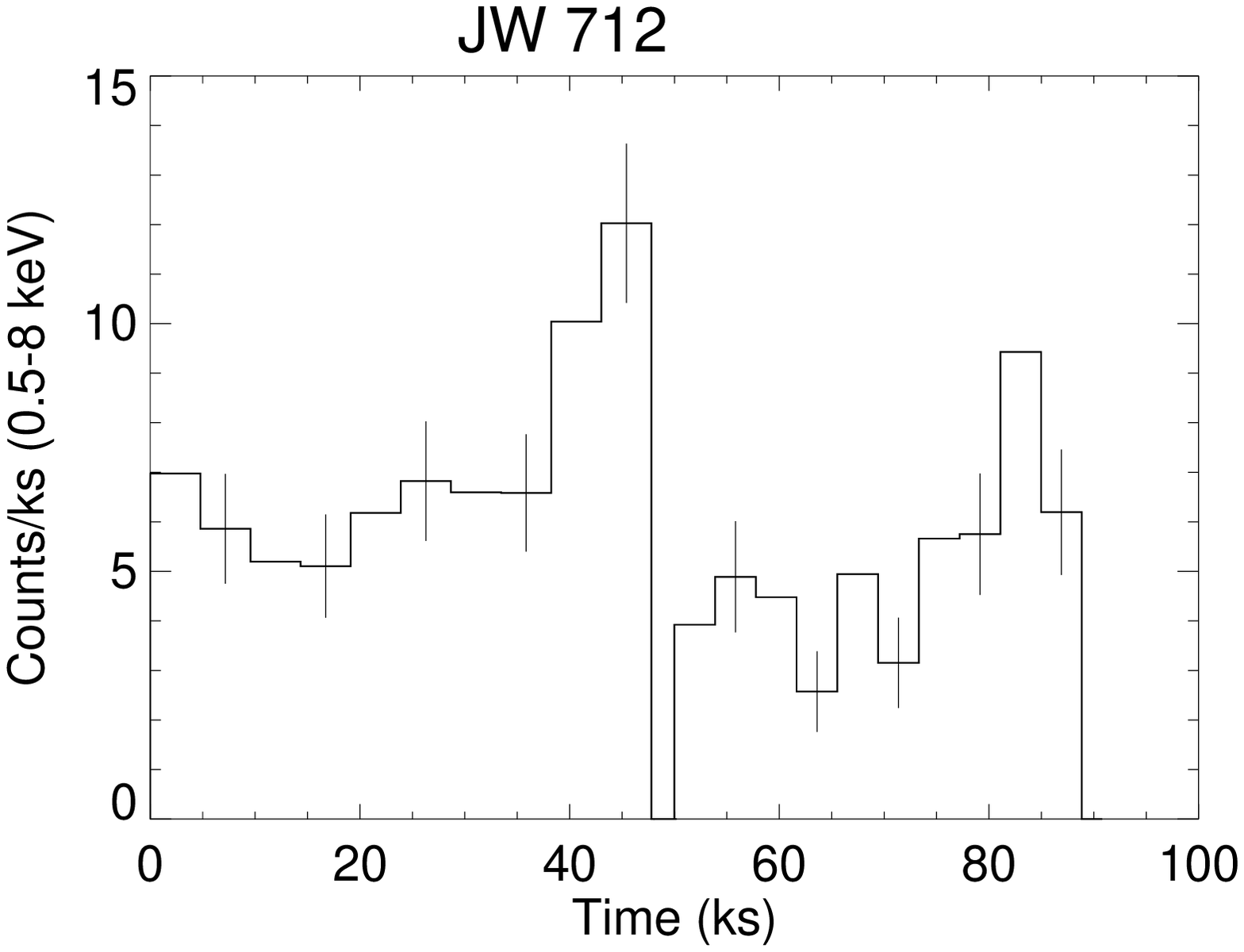}
\includegraphics[scale=0.30]{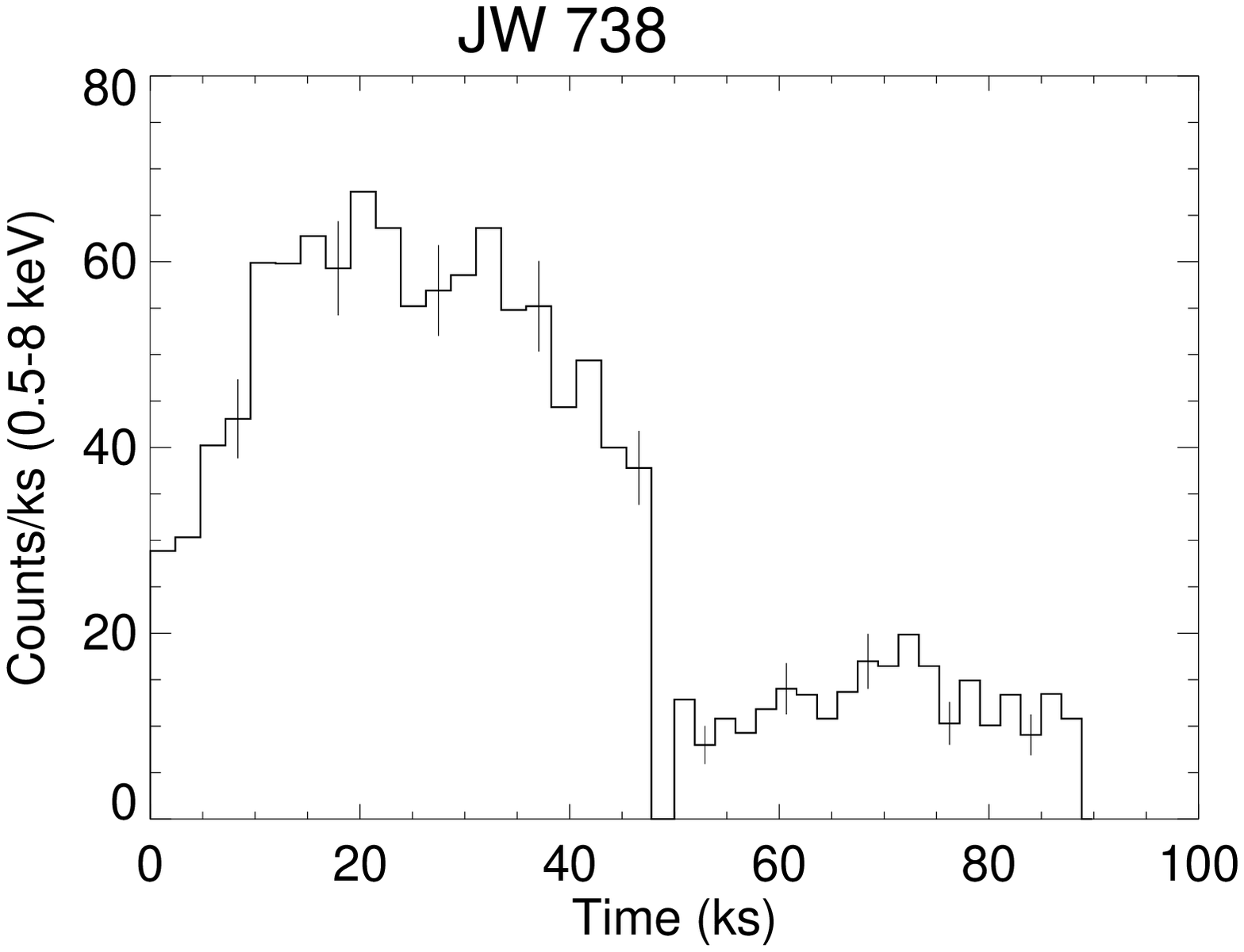}
  \end{minipage} \\ [0.3in]
  \begin{minipage}[t]{1.0\textwidth}
  \centering
\includegraphics[scale=0.30]{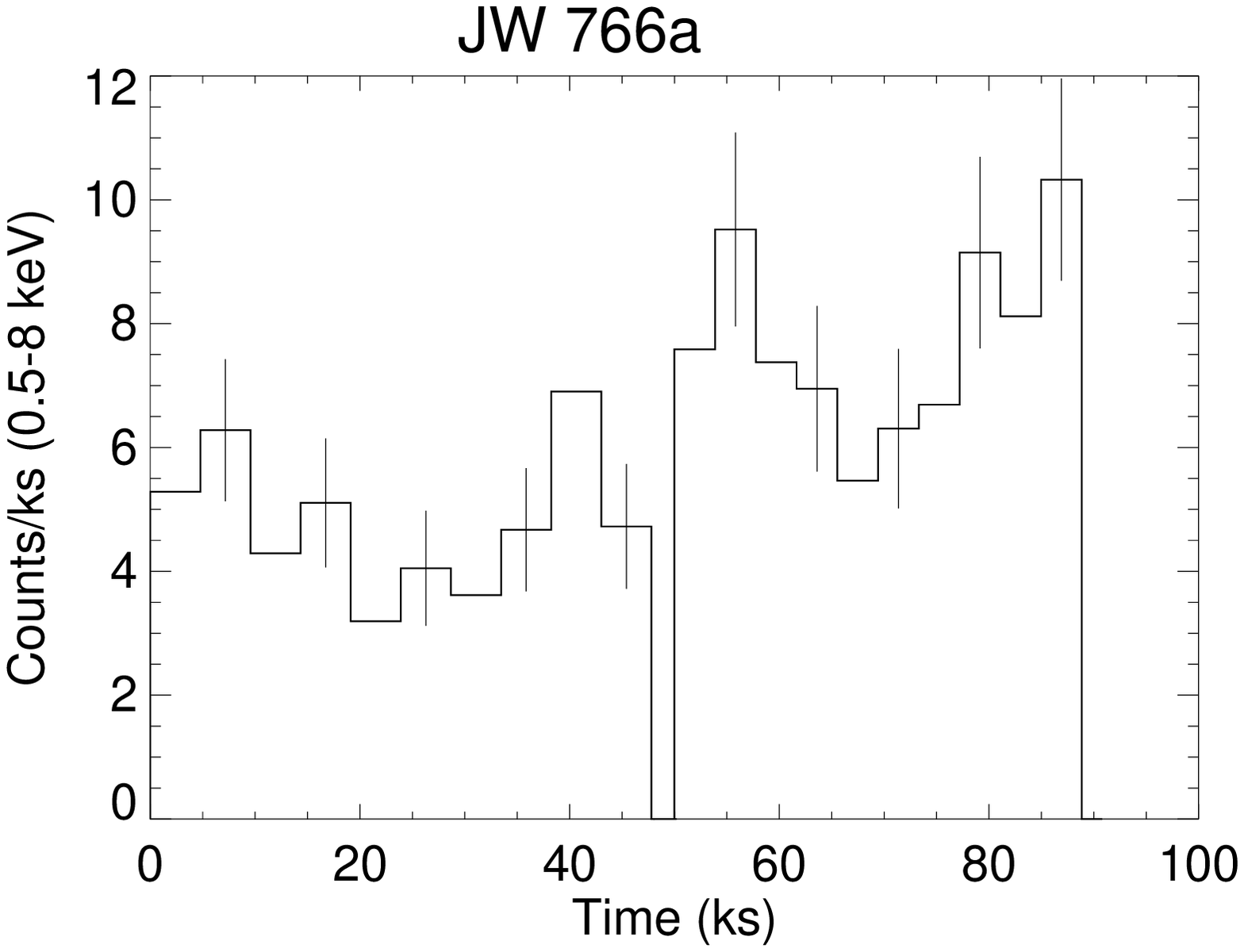}
\includegraphics[scale=0.30]{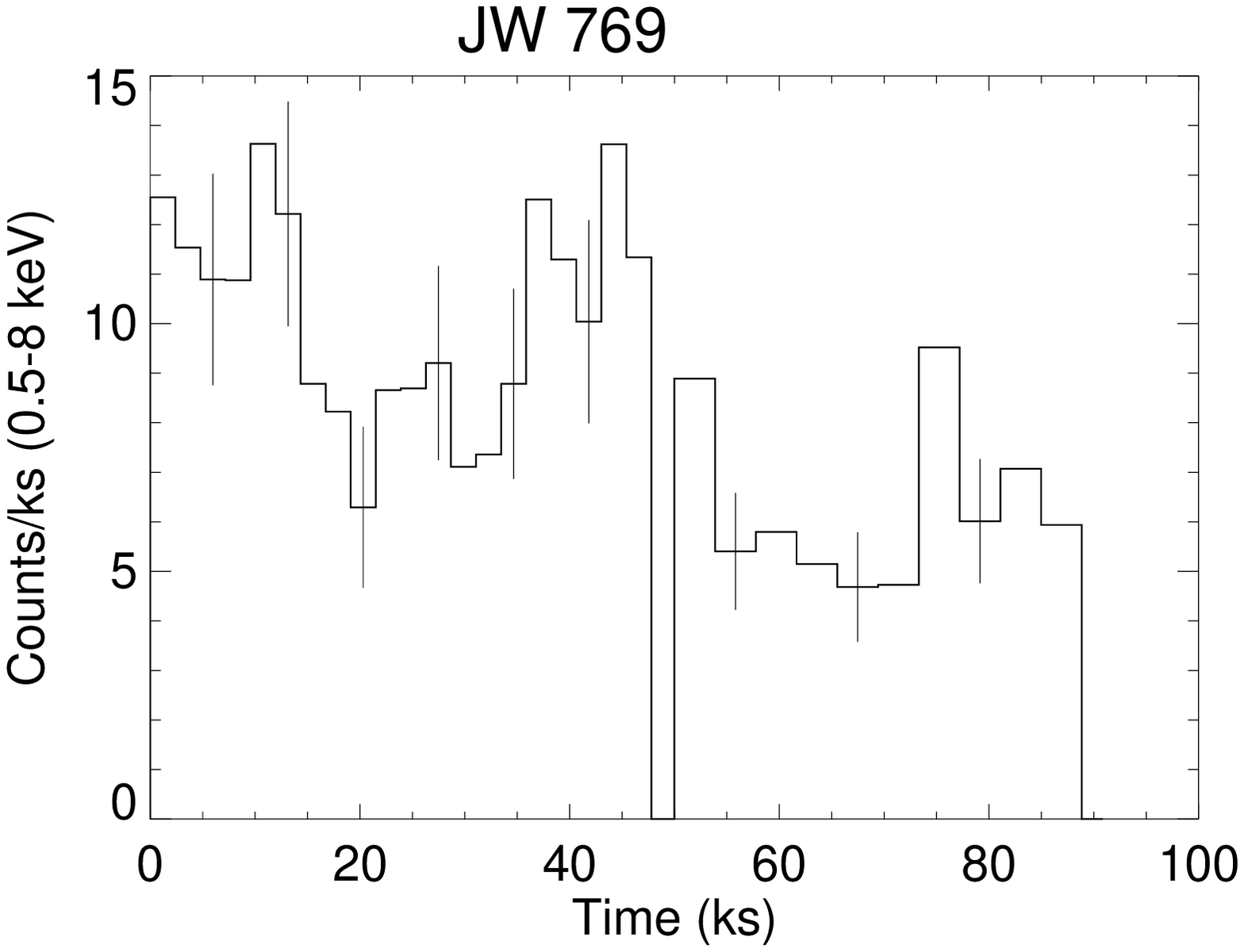}
\includegraphics[scale=0.30]{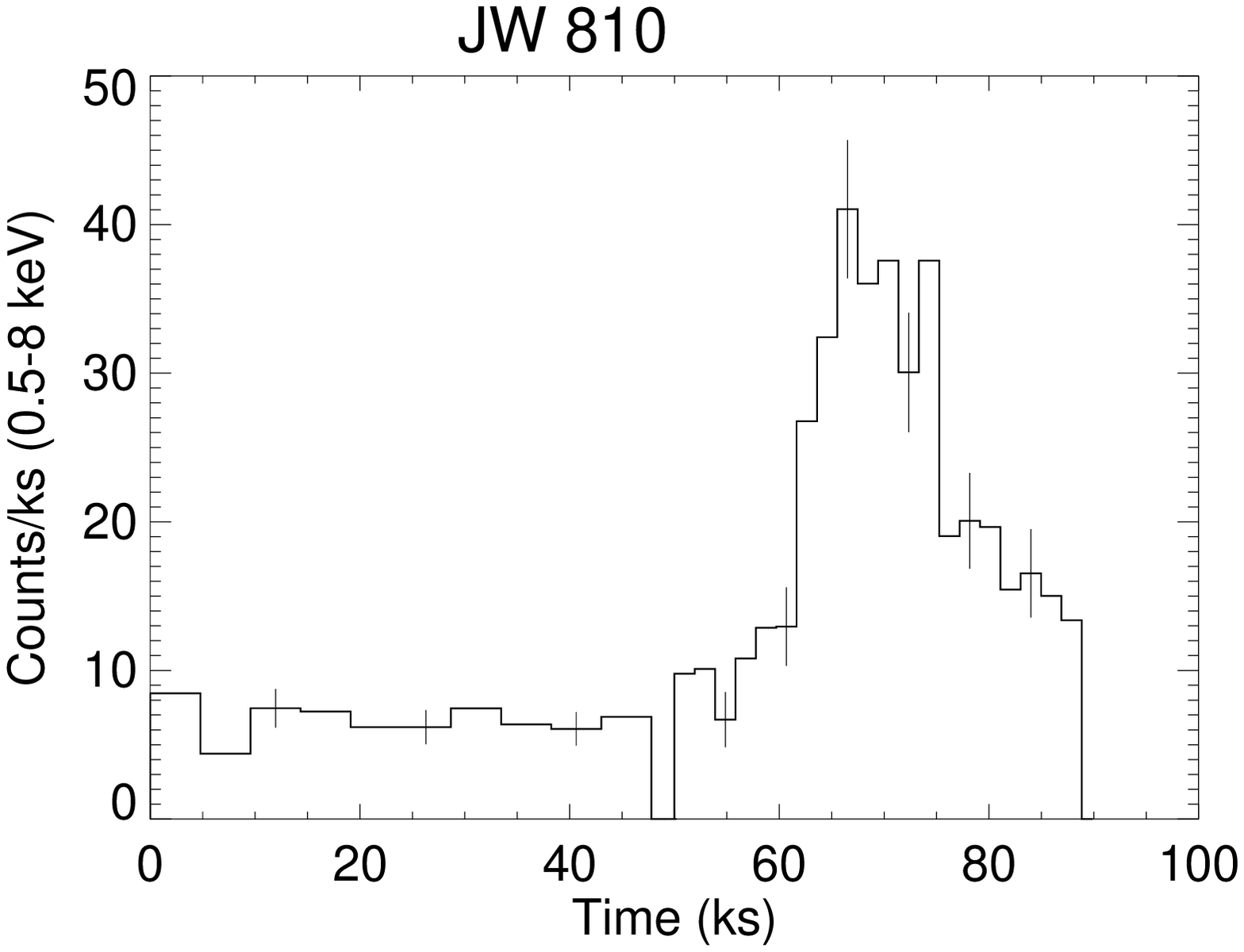}
  \end{minipage} \\ [0.3in]
  \begin{minipage}[t]{1.0\textwidth}
  \centering
\includegraphics[scale=0.30]{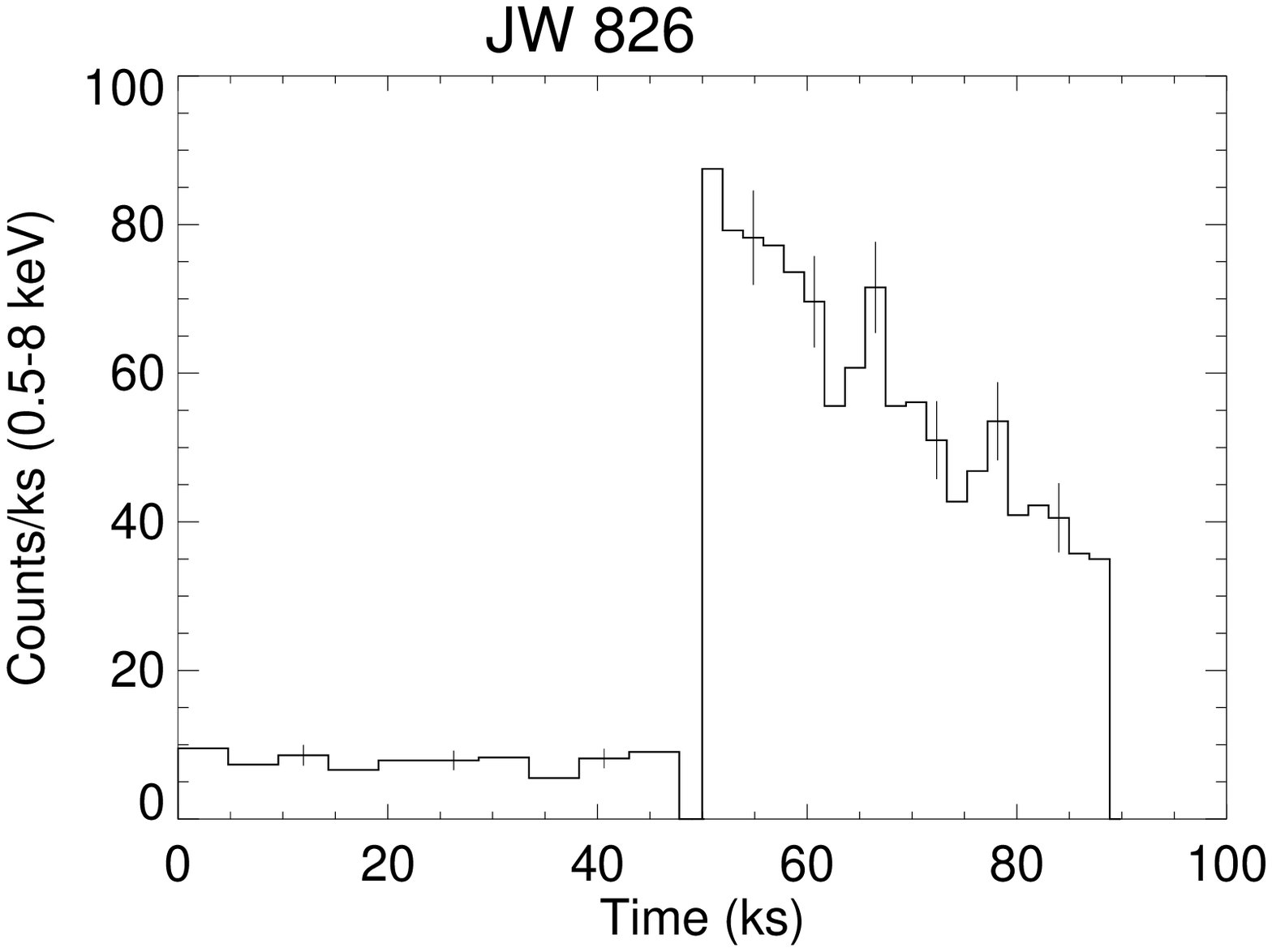}
\includegraphics[scale=0.30]{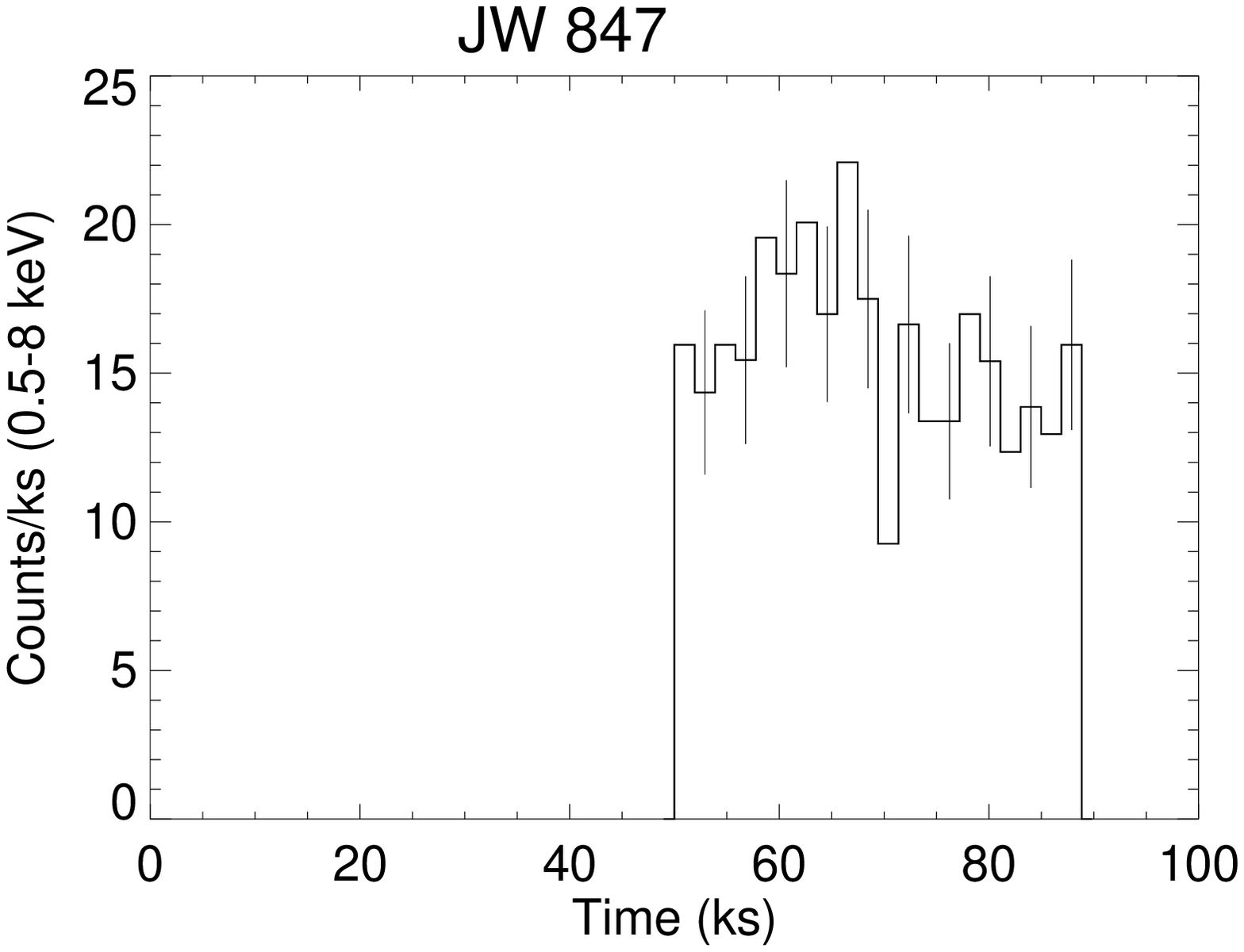}
\includegraphics[scale=0.30]{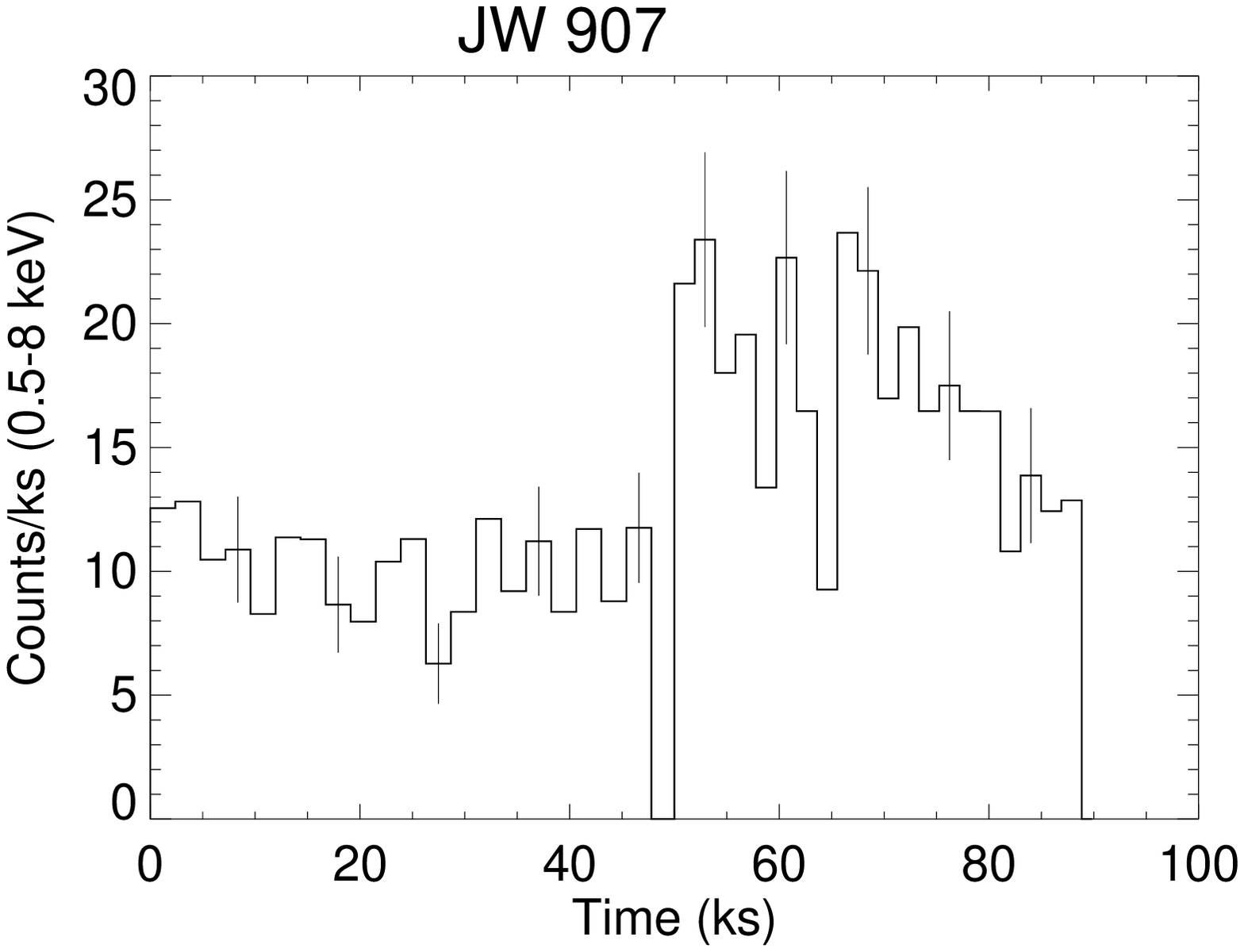}
   \end{minipage}
\end{figure}

\end{document}